\documentclass{article}

\usepackage[english]{babel}

\usepackage[letterpaper,top=2cm,bottom=2cm,left=2cm,right=2cm,marginparwidth=1.75cm,margin=1in]{geometry}
\usepackage{amsmath}
\usepackage{graphicx}
\usepackage[colorlinks=true, allcolors=blue]{hyperref}
\usepackage{soul}
\usepackage{xcolor}
\usepackage{authblk}

\usepackage{subcaption} 
\usepackage{booktabs}
\usepackage[font=small,labelfont=bf]{caption}
\usepackage[margin=1in]{geometry}
\usepackage{hyperref}
\hypersetup{colorlinks=true,linkcolor=black,citecolor=blue}
\usepackage{tabularx}
\usepackage{multirow}
\usepackage{array}
\usepackage{xcolor}
\usepackage[section]{placeins} 

\title{Length and Sequence-Dependent Dipolar and Electronic Properties of $\alpha$-helical Polypeptides}
\author[1,2]{Faezeh Shabani}
\author[3]{Florian Günther}
\author[1,2*]{Sibylle Gemming}

\affil[1]{Institute of Physics, Chemnitz University of Technology, Chemnitz, Germany}
\affil[2]{Research Center for Materials, Architectures and Integration of Nanostructures (MAIN), Chemnitz University of Technology, Chemnitz, Germany}
\affil[3]{ Departamento de Física Instituto de Geociências e Ciências Exatas, Universidade Estadual Paulista,Rio Claro, SP, Brazil}

\affil[*]{Corresponding author: \texttt{sibylle.gemming@physik.tu-chemnitz.de}}

\date{}
\usepackage{subcaption}
\begin{document}
\maketitle

\begin{abstract}

As a class of sequence-defined biopolymers, $\alpha$-helical polypeptides combine the chemical tunability of synthetic macromolecules with the highly ordered backbone architecture of natural proteins. Here, density-functional tight-binding calculations are used to determine how chain length and sequence chemistry govern the dipolar and electronic properties of three $\alpha$-helical polypeptide families: polyalanine [AAAAA]$_n$, leucine-modified [AAAAL]$_n$, and cysteine-capped, lysine-containing C[AAAAK]$_n$. 
Increasing chain length produces systematic shifts and splitting of the molecular orbital levels together with an approximately additive increase in the molecular dipole. Analysis of the C[AAAAK]$_3$ helix shows that aligned backbone carbonyl groups provide the dominant contribution to the total dipole, supporting a dominant backbone contribution to the macrodipole, whose growing electrostatic potential is consistent with an intrinsic Stark-like contribution to the length-dependent orbital-energy evolution. Side-chain and terminal modifications introduce localized perturbations to frontier electronic states while preserving this global backbone-driven trend. Residue-resolved frontier levels additionally form a staircase-like energy landscape along the helical scaffold: individual peptide units retain large local HOMO–LUMO gaps of approximately 5 eV, whereas spatially proximate hydrogen-bonded residues exhibit smaller inter-residue energy separations. Application of external electrostatic fields modifies the frontier-level separation in a direction-dependent manner and induces structural distortions predominantly near the flexible termini, while the helical core remains comparatively robust. By establishing these fundamental sequence–structure–property relationships, this work provides a foundational blueprint for the rational design of peptide-based biomacromolecules tailored for next-generation bio-electronic interfaces, biosensors, and artificial electron-transfer systems.
\end{abstract}

\section{Introduction}

$\alpha$-helical peptides have emerged as versatile building blocks for biological and biomimetic functional materials because they combine a structurally persistent hydrogen-bonded scaffold with sequence-level chemical programmability and sensitivity to their molecular environment \cite{BEESLEY2019175,levin_biomimetic_2020}. Their ordered architectures support functions ranging from molecular recognition and self-assembly to charge transfer, molecular electronics, and field-responsive behavior \cite{lee_long_1992,https://doi.org/10.1002/cplu.201500121,han_advances_2025,Slawig2020}. This combination of structural resilience and chemical adaptability makes the $\alpha$-helix an attractive platform for constructing sequence-defined functional materials. First described by Pauling, Corey, and Branson in 1951, the $\alpha$-helix is also distinguished by a collective electrostatic property: the vectorial alignment of individual peptide-bond dipoles, each on the order of 3--4~D, generates a substantial macrodipole along the helical axis \cite{Pauling1951,Hol1985,Sali1988}. High-resolution crystallographic analyses further indicate that peptide bonds within $\alpha$-helices display a subtly enhanced enol-like character compared with those in $\beta$-strands, providing a more detailed picture of peptide-bond electronic structure within the helical environment \cite{Panjikar2025}. The resulting collective dipole gives rise to a substantial internal electric field that influences protein function, including active-site electrostatics, ligand binding, and the $pK_a$ shifts of terminal residues \cite{Hol1985,Scholtz1992}.

The magnitude of the helix dipole depends on structural variables including chain length, terminal functionalization, and side-chain chemistry \cite{Sengupta2005,Marqusee1987,Groebke1996,Ponomarev2012}. More generally, synthetic polypeptides provide sequence-defined macromolecular architectures in which the number and chemical identity of repeating units regulate secondary structure and collective molecular properties \cite{10.1021/acs.biomac.7b00534,li_helix-specific_2023}. The programmability of $\alpha$-helical architectures is particularly well established in coiled-coil systems, where sequence--structure relationships enable rational control over stability, molecular recognition, oligomerization, and higher-order assembly, making such motifs increasingly useful as modular platforms for biomolecular and functional-material design \cite{https://doi.org/10.1002/cbic.70585}. Molecular length can likewise serve as a functional design parameter in $\alpha$-helical polypeptide systems \cite{10.1021/acs.biomac.4c01716}; notably, increasing helical length has recently been shown to amplify the macrodipole and thereby enhance ionic conductivity and electrochemical stability in peptide-based materials \cite{chen_helical_2024}. These characteristics make $\alpha$-helical polypeptides useful model systems for understanding how local dipolar units and sequence-dependent interactions give rise to collective properties of a polymer chain.

The functional properties of helical peptides have consequently been rationalized largely at the collective structural and electrostatic level, through the interplay of backbone hydrogen bonding, residue-specific conformational preferences, side-chain interactions, and long-range electrostatics \cite{Hol1985,https://doi.org/10.1002/pro.5560030514,https://doi.org/10.1002/1097-0282(2001)60:5<366::AID-BIP10177>3.0.CO;2-5,errington_structure_2006}. Within this framework, helix geometry and dipolar ordering provide intuitive structure--property relationships linking molecular architecture to stability, polarization, self-assembly, and response to environmental perturbations \cite{Hol1985,li_helix-specific_2023}. Such principles help explain why $\alpha$-helical scaffolds can combine structural persistence with substantial chemical tunability.

The favorable properties of helical peptides have therefore largely been rationalized at the collective structural level, through the interplay of steric organization, backbone hydrogen bonding, sequence-dependent interactions, and long-range electrostatics. At this mesoscopic level, helix geometry and dipolar ordering provide intuitive structure–property relationships linking molecular architecture to stability, polarization, assembly, and response to external perturbations. Such descriptions have been highly successful for understanding why $\alpha$-helical scaffolds are simultaneously structurally persistent and chemically tunable.

An important consequence of this ordered arrangement is the development of an electrostatic potential gradient along the helix. As the number of similarly oriented peptide-bond dipoles increases, their approximately additive contribution increases the molecular macrodipole and changes the electrostatic environment experienced by electronic states located at different positions along the chain \cite{Hol1985,Sengupta2005}. In molecular systems constructed from polar repeating units, analogous internally generated potential gradients have been shown to shift frontier orbital energies, localize occupied and unoccupied states toward opposite molecular regions, and reduce the frontier-state gap in a manner analogous to an externally induced Stark effect \cite{Rissner2011}. We therefore refer to this mechanism as a macrodipole-induced or intrinsic Stark-like effect. A pivotal recent experimental and theoretical study of an $\alpha$-helical peptide scaffold demonstrated that the intrinsic helix dipole field can directly modulate proton-coupled electron-transfer kinetics, further establishing a structure–field–function relationship associated with the $\alpha$-helical macrodipole \cite{10.1021/jacs.6c06120}. These observations suggest that chain length in an $\alpha$-helical polypeptide may influence not only its total dipole moment but also the energetic and spatial organization of its electronic states.

 While the peptide backbone provides the repeated polar units responsible for the global macrodipole, the electronic structure can additionally be modified by the chemistry of individual side chains and terminal groups. Side-chain electrostatics, particularly from basic residues such as lysine, interact with the helix dipole in a position-dependent manner and can modify local charge distributions and conformational stability \cite{Groebke1996,Ponomarev2012}. Terminal functionalization can likewise enhance or screen the electrostatic potential generated by the backbone \cite{Sengupta2005,Marqusee1987}. These two contributions need not affect the electronic structure in the same manner: the repeated backbone can determine the global chain-length-dependent trend, whereas chemically distinct side chains or terminal groups may introduce localized states or perturb particular frontier orbital energies. Distinguishing these global backbone effects from local sequence-dependent contributions is therefore important for developing sequence–property relationships in electronically functional polypeptides.

The coupling between molecular architecture, electrostatics, and electronic structure is particularly relevant to the growing use of $\alpha$-helical peptides as functional molecular materials. Helical peptides have been investigated as mediators of long-range electron transfer and as molecular bridges in electronic junctions \cite{lee_long_1992,https://doi.org/10.1002/cplu.201500121,han_advances_2025,chen2022modulation}, and computational modeling continues to guide the rational design of natural peptide sequences into functional electronic conductors and nanowires~\cite{10.1021/bm900524v}. Single-molecule measurements on $\alpha$-helical polyalanine derivatives have demonstrated length-dependent tunneling conductance and substantial sensitivity of the transport barrier to applied electric fields \cite{Slawig2020}, while theoretical and experimental studies indicate that peptide secondary structure and electronic coupling can strongly influence charge-transfer and transport behavior \cite{Grozema2010, Brisendine2018,Samajdar2024}. The intrinsic dipolar environment of $\alpha$-helices may therefore provide an additional molecular design parameter for controlling electronic processes. Chiral $\alpha$-helical peptides have also become model systems for studies of chiral-induced spin selectivity (CISS), where spin-dependent transmission has been associated with molecular chirality, electrostatic potential gradients, molecular orientation, and molecule–electrode coupling \cite{Naaman2012,Bloom2024,EckshtainLevi2016,Nguyen2024}.Complementary analytical models of spin--orbit-coupled particles under helical confinement have shown that helical geometry can produce Rashba-type spin splitting while preserving time-reversal symmetry, providing an idealized framework for distinguishing geometry-induced spin--orbit effects from the additional mechanisms required for CISS \cite{Muresan2026}. Although a quantitative description of charge or spin transport requires explicit transport and spin-dependent calculations, these observations motivate a detailed understanding of the underlying electronic-state organization and its response to intrinsic and externally applied electrostatic fields.

Despite this extensive understanding at the structural and electrostatic levels, a systematic molecular-level picture connecting these collective properties to the underlying electronic structure remains incomplete \cite{improta_structure_2001,marianski_aqueous_2011}. A refined electronic description of realistic helical peptides is challenging because the relevant systems rapidly reach hundreds of atoms, for which conventional first-principles calculations become increasingly demanding \cite{cole_applications_2016}. Moreover, their electronic behavior reflects coupled phenomena across multiple structural scales, spanning local peptide bonds and side-chain states, residue-to-residue interactions, hydrogen-bonded turns, and the collective macrodipole of the entire helix. Systematically resolving these connections therefore requires electronic-structure methods that retain chemically resolved orbital information while remaining tractable across multiple sequences, chain lengths, and external perturbations. Recently, we employed self-consistent-charge density-functional tight-binding (SCC-DFTB) at this intermediate level of description to derive effective interaction potentials for the self-assembly of polyalanine $\alpha$-helices, providing an atomistically resolved description of interhelical packing and supramolecular organization \cite{10.1021/acsami.6c01087}. In that treatment, however, the helices were represented as ideal infinite periodic structures, deliberately excluding terminal groups and suppressing global-dipole interactions in order to isolate local intermolecular interactions. The electronic consequences of finite chain length, molecular termination, and the emergence of the helix macrodipole therefore remain outside the scope of such models.
Consequently, a systematic electronic-structure analysis that directly links side-chain substitution, helix length, and macrodipole formation to the evolution of frontier orbitals in well-defined oligomeric helices remains limited. In particular, three related questions remain insufficiently resolved: (i) how the cumulative dipoles of individual peptide units translate into systematic changes in molecular orbital energies as the $\alpha$-helical chain grows; (ii) which features of this electronic response originate from the common peptide backbone and which are locally modified by sequence and terminal chemistry; and (iii) how an externally applied electrostatic field interacts with the intrinsic macrodipole to alter the electronic and structural response of the helix.

In this work, we build upon our previous work  \cite{10.1021/acsami.6c01087}
and address these questions again using SCC-DFTB calculations for three 
$\alpha$-helical polypeptide families with increasingly complex side chain functionalities: polyalanine [AAAAA]$_n$, still unpolar aliphatic, but sterically more demanding leucine-modified [AAAAL]$_n$, and the experimentally studied, N-terminally cysteine-capped, lysine-containing C[AAAAK]$_n$ with polar end groups. By varying chain length across the three families, we relate the evolution of molecular orbital energies to the growth of the helix macro dipole and quantify the contribution of aligned backbone carbonyl groups to this collective polarization. Residue- and functional-group-resolved electronic analyses are then used to distinguish the global backbone-controlled response from local sequence-dependent perturbations and to identify the staircase-like organization of frontier-state energies along the helical scaffold. Finally, externally applied electrostatic fields are used to probe the coupling between the intrinsic dipolar architecture, frontier-level energetics, and structural flexibility. Together, these calculations establish a molecular-level structure–property relationship connecting chain length, sequence chemistry, and $\alpha$-helical dipolar ordering to the emergent electronic properties of sequence-defined polypeptides, while providing a basis for understanding their relevance to charge-transfer and field-responsive biomolecular materials.

\begin{figure}[htbp]
    \centering
    \includegraphics[width=0.7\textwidth]{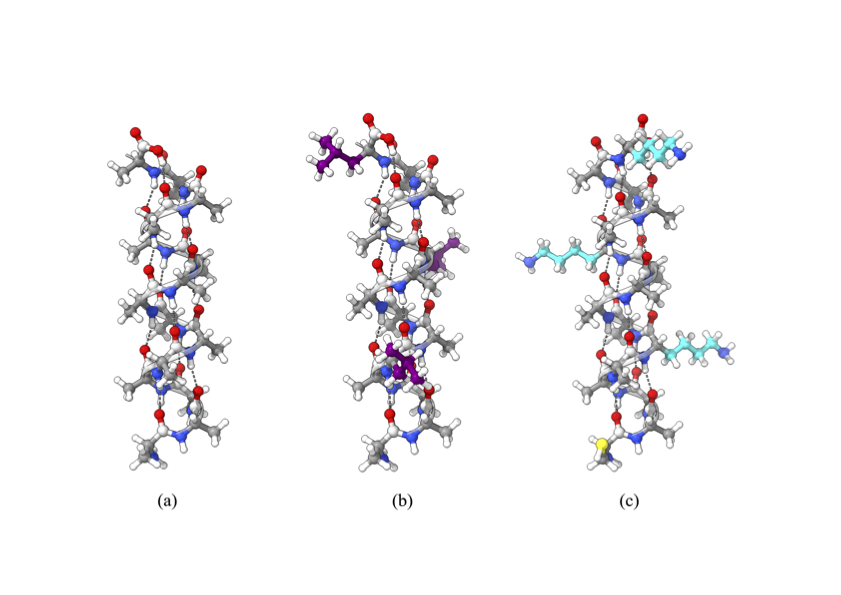}
    \caption{$\alpha$-helix chains under study (a) Polyalanine, (b) [AAAAL]$_{n}$, and (c) C[AAAAK]$_{n}$.}
    \label{fig:1}
\end{figure}

\section{Methods and Computational Details}

The initial structural configurations of the three groups of $\alpha$-helices were generated using the Avogadro molecular modeling software\cite{hanwell_avogadro_2012}. These models were constructed to investigate the intrinsic electronic properties of isolated secondary structures of polypeptides in the gas phase and treated as non-periodic cluster models. 

In detail, to investigate how the nature of the side chains in  $\alpha$-helix polypeptides influences the electronic properties, we modeled the following three groups of polyalanine-based helices: 
The first group, illustrated in Figure~\ref{fig:1}a, consists of pure polyalanine $\alpha$-helices, starting with 5 residues and progressively extending in steps of 5 up to 35 residues. As the number of residues increases, the characteristic $\alpha$-helical secondary structure becomes more pronounced due to backbone hydrogen bonding and steric constraints \cite{doi:10.1073/pnas.96.25.14258}. However, experimentally, modifications of the purely alanine-based amino acid sequence have been employed to enhance the solubility of the $\alpha$-helices in various media and to anchor the molecule on surfaces and scanning tunneling microscopy tips \cite{Slawig2020}. Thus, two further sets of molecules were chosen which introduce also the influence of the side chain modifications in a step-wise manner:
The second group, shown in Figure~\ref{fig:1}b, introduces a periodic perturbation into the polyalanine sequence by inserting a leucine residue after every four alanines, yielding sequences of the form [AAAAL]$_{n}$ with $n = 1,\dots,7$. 
This set of purely aliphatic polypeptides is a model system, which is intended to isolate only the sterical aspect of the experimentally used side chain modifications.
Finally, in the third group (Figure~\ref{fig:1}c), the leucine residues were replaced by lysine and a cysteine residue is added at the N-terminus, resulting in C[AAAAK]$_{n}$, with n=1,\dots,7. The N-terminal cysteine provides a potential anchoring site for interaction with metal surfaces such as gold, whereas the lysine substitutions introduce chemically distinct side chains into the repeating sequence.

Electronic structure calculations were performed using the DFTB+ package (release 24.1)\cite{hourahine_dftb_2020}. We employed the Self-Consistent Charge (SCC-DFTB) formalism, which approximates the Kohn–Sham density functional theory (DFT) total energy. 
Atomic interactions and electronic integrals were described using the mio-1-1 Slater–Koster parameter set 
\cite{PhysRevB.58.7260,Perdew1996,boleininger_gaussian_2016}. 
To account for dispersion interactions, the SCC-DFTB calculations were supplemented with a Lennard-Jones dispersion correction using parameters from the Universal Force Field (UFF).

For all systems, the SCC cycles were carried out with a convergence tolerance of $10^{-5}$ atomic units to ensure a well-converged electronic density. Full geometry optimizations were performed to relax the structures to their local minima. The optimization process was considered converged when the maximum absolute gradient element reached a threshold of $10^{-4}$ Hartree/Bohr. 

Under these conditions, the standard SCC-DFTB method provides an appropriate balance between computational efficiency and the accurate description of fundamental electronic trends\cite{https://doi.org/10.1002/jcc.24866}. 
SCC-DFTB has previously been benchmarked for glycine- and alanine-based peptide models and shown to reproduce the geometries and relative energies of helical and other secondary-structure motifs in good agreement with higher-level DFT calculations \cite{elstner_dft_2000, ELSTNER2001203}. The isolated chain model employed here was chosen specifically to isolate the intrinsic electronic and dipolar contributions of the $\alpha$-helical backbone from external screening effects \cite{Sengupta2005}. This approach allows us to establish the fundamental structure-property relationships inherent to the polypeptide chain itself—a necessary baseline for understanding how these properties are modulated by environmental factors. We note that the effective dipole moment in solution can differ significantly from the intrinsic gas-phase value, as the interaction with the solvent can either screen the dipole (in a uniform high-dielectric medium) or, in certain implicit solvent models like COSMO, generate a reaction field that reinforces it\cite{PICHIERRI2003197}. 
Previous studies on small helical peptides indicate that dispersion has, in general, little effect on geometries and relative energetics for compact helical structures \cite{10.1021/ct600352e}. To assess the influence of the dispersion treatment on the systems investigated here, comparative calculations were performed for selected representative helical structures using three approaches: the Lennard-Jones correction with UFF parameters, the D3 dispersion correction, and no dispersion correction. The results of these comparisons are presented in the Supporting Information(SI) and allow the sensitivity of the calculated properties to the dispersion treatment to be evaluated for the selected structures.

Following structural relaxation, the electronic properties, including the energy levels of the molecular orbitals (MOs) and the Projected Density of States (PDOS), were analyzed. The PDOS was calculated by applying a Gaussian broadening with the broadening factor of 0.1 eV to the discrete energy levels of the molecular orbitals, projected onto specific fragments such as individual amino acids and atomic species. The Mulliken charges were used to calculate the total dipole moment of the helix as well as the dipole moments of specific fragments. Furthermore, to visualize the spatial distributions of the highest occupied molecular orbitals (HOMO) and lowest unoccupied molecular orbitals (LUMO), the \texttt{waveplot} tool within the DFTB+ package was employed. Finally, the principal helix axis was determined using principal component analysis (PCA) of the C$\alpha$ atomic coordinates, with the first principal component defining the helix axis for structural orientation and alignment.

\section{Result and discussion}
To establish how the molecular structure and sequence of $\alpha$-helical peptides influence their electronic properties, we first examine the evolution of the frontier electronic states with peptide length and side-chain composition. We then analyze the spatial and residue-resolved character of these states and relate the resulting energetic gradients to the intrinsic macrodipole of the $\alpha$-helix. Finally, we investigate how this intrinsic electrostatic landscape responds to an externally applied electrostatic perturbation. Together, these analyses provide a consistent picture of the interplay between peptide structure, molecular polarity, and frontier electronic structure.
\subsection{Chain-Length and Side-Chain Dependence of the Frontier Electronic Structure}
The three groups of model structures allow distinguishing the influence of steric and electronic side chain modifications on the evolution of the orbital energy spectrum with varying length of the $\alpha$-Helix.

\begin{figure}[htbp]
    \centering
    \begin{subfigure}[b]{0.32\textwidth}
        \centering
        \includegraphics[width=\textwidth]{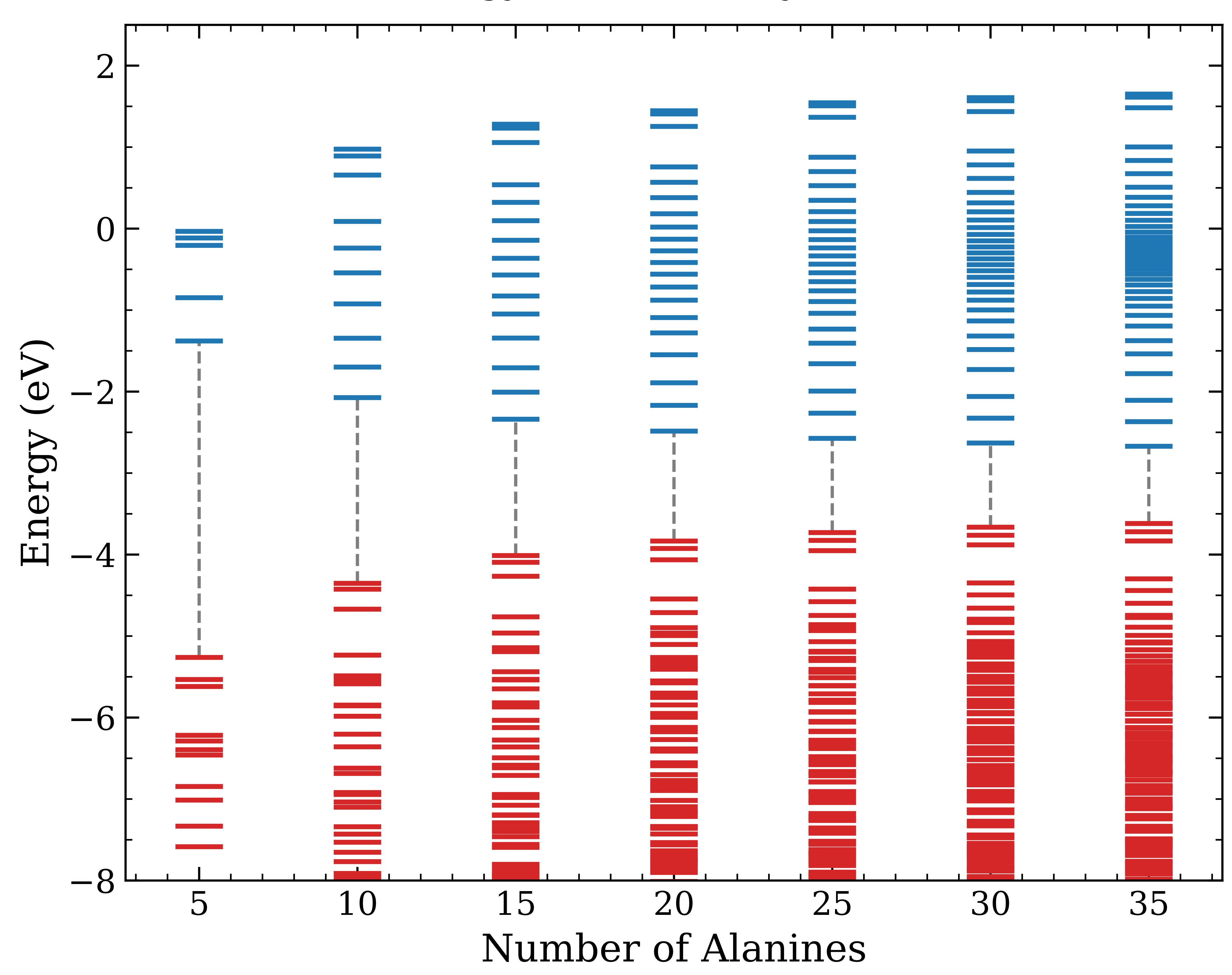}
        \caption{}
        \label{fig:2a}
    \end{subfigure}
    \hfill
    \begin{subfigure}[b]{0.32\textwidth}
        \centering
        \includegraphics[width=\textwidth]{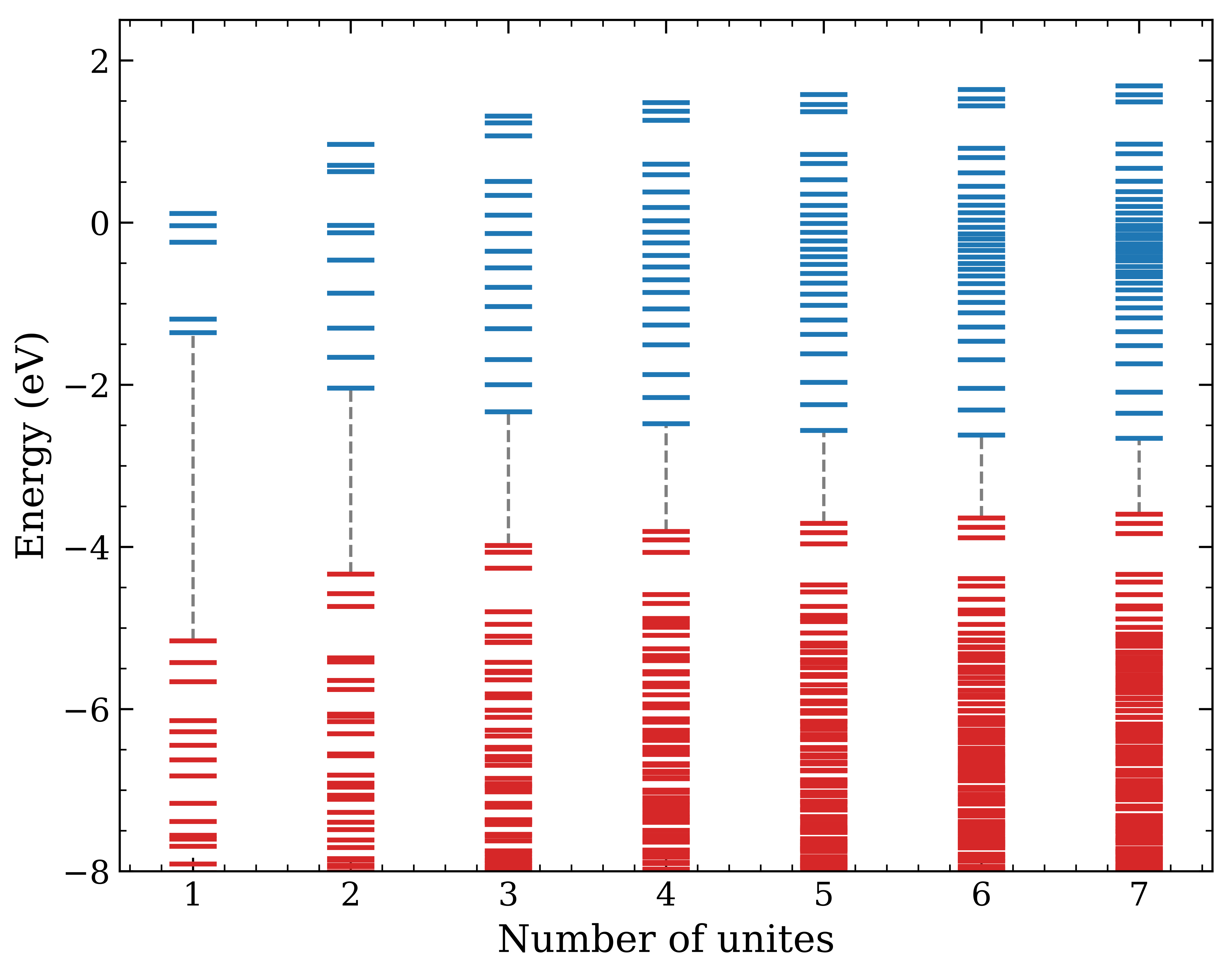}
        \caption{}
        \label{fig:2b}
    \end{subfigure}
    \hfill
    \begin{subfigure}[b]{0.32\textwidth}
        \centering
        \includegraphics[width=\textwidth]{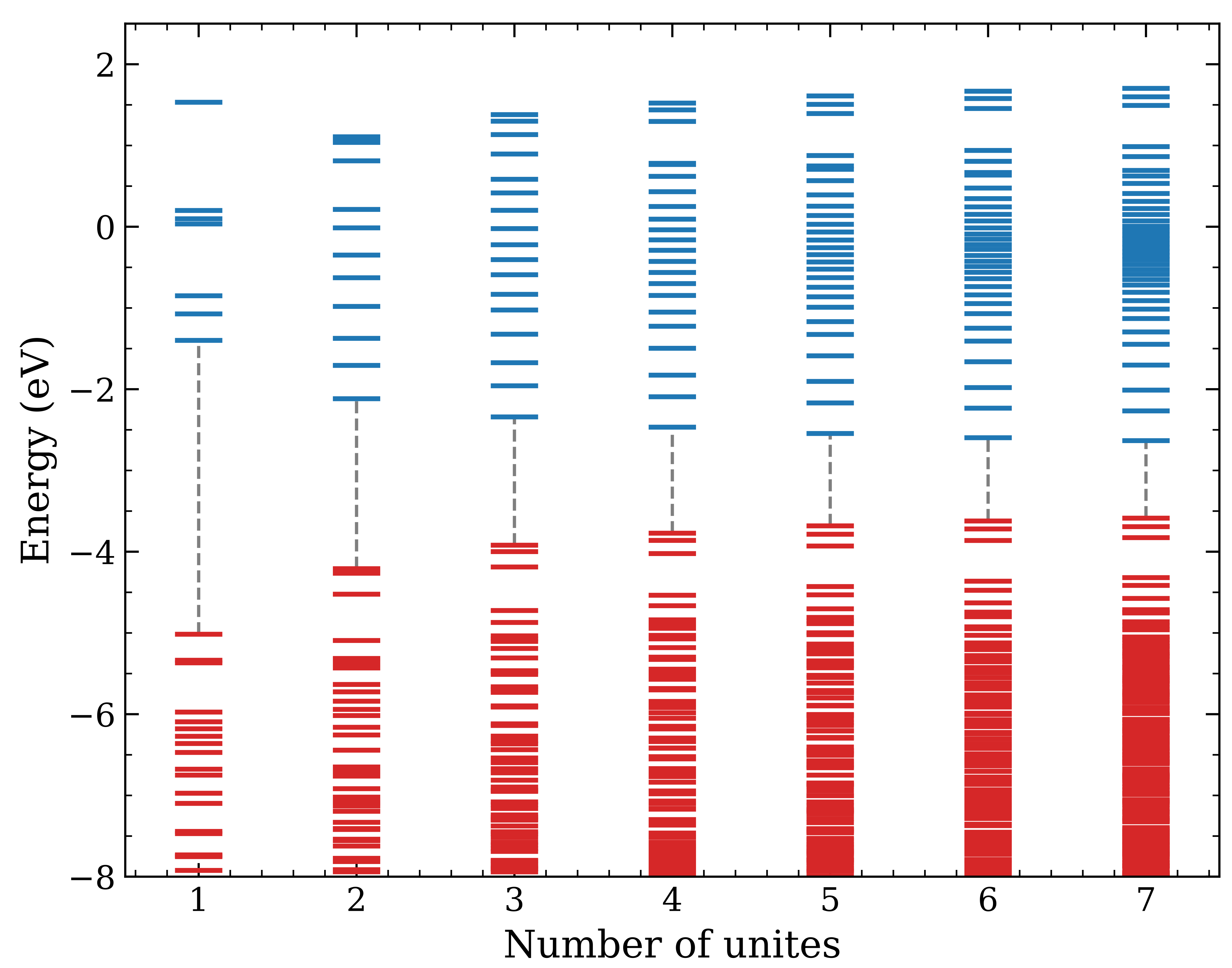}
        \caption{}
        \label{fig:2c}
    \end{subfigure}
    \caption{Energy levels of the molecular orbitals near the Fermi level for (a) Polyalanine, (b) [AAAAL]$_{n}$, and (c) C[AAAAK]$_{n}$ as a function of chain length.}
    \label{fig:2_all}
\end{figure}

Figure~\ref{fig:2a} shows the energy levels of the MOs of the three studied $\alpha$-helical peptides in the vicinity of the HOMO and LUMO levels as a function of the chain length, given in numbers of alanine residues (a) or of [AAAAX] units with X = L in (b) and X = K in (c). The unoccupied orbitals are represented in blue and the occupied ones are shown in red. Increasing the number of residues increases the number of molecular states and leads to systematic shifts in their energies. Consequently, the occupied and unoccupied levels become progressively more closely spaced as the helix length increases. For the finite oligomers considered here, these states remain discrete; however, their increasing density with chain length is consistent with the development of increasingly dense electronic-state manifolds as the polymer limit is approached (see SI). At the same time, the HOMO and LUMO shift systematically, resulting in a progressive narrowing of the HOMO–LUMO (H–L) gap.

The energy diagram of [AAAAL]$_{n}$, presented in Figure~\ref{fig:2b}, exhibits the same overall splitting behavior and the converging trend of the HOMO--LUMO gap. For shorter chains, particularly those containing one to three repeating units, the introduction of leucine modifies the splitting and spacing of both occupied and unoccupied energy levels compared with pure polyalanine. However, these differences become less pronounced as the number of repeating units increases, and the overall distribution of the energy levels approaches that of the corresponding polyalanine helices. Thus, periodic substitution of alanine by the larger but still aliphatic leucine side chain does not qualitatively alter the characteristic length dependence of the frontier electronic structure.

A similar trend is observed for the third group of helices, C[AAAAK]$_{n}$, shown in Figure~\ref{fig:2c}, which displays the same progressive energy level splitting and narrowing of the H--L gap with increasing chain length as observed for polyalanine and [AAAAL]$_{n}$. In these helices, cysteine is present at the N-terminus in addition to the substitution of lysine for leucine in the repeating sequence. For shorter chains, these chemical modifications lead to changes in the splitting of the energy levels and the appearance of two additional states in the unoccupied region compared with the purely aliphatic systems. As the chain length increases, the differences in the overall level distribution become less pronounced, although the chemically distinct residues continue to contribute to specific molecular states. In particular, the sulfur-containing group of the cysteine residue contributes to higher-energy unoccupied states, while the additional lone-pair states of the NH$_{\mathrm{2}}$ groups of the lysine residues are located within the manifold of occupied orbitals (see SI for more details)., these chemically distinct residues, introduce additional states into both the occupied and unoccupied regions of the spectrum. In particular, the sulfur-containing group of the cysteine residue contributes to higher-energy unoccupied states. The additional lone pair states of the NH$_{\mathrm{2}}$ groups of the lysine residues are located within the manifold of occupied orbitals (see SI for more details).
Despite the differences in the detailed energy-level distributions, particularly for shorter chains, the three helical systems exhibit similar length-dependent behavior of the HOMO and LUMO energies and a progressive narrowing of the fundamental H--L gap. This indicates that although side-chain composition influences individual molecular states, the overall evolution of the frontier electronic structure with increasing helix length is largely preserved across the three sequences.

In accordance with  ref. \cite{10.1021/jp025556i} the H--L gap shrinks with increasing chain length. However, with increasing chain length the H--L band gap of the single polyalanine chain neither exhibits the typical particle-in-the-box-type inverse proportionality on the number of residues, nor does it just lead to an increased density of electronic states around the monomer levels. It is, however reminiscent of the gradual reduction of the H--L gap, which oligo-thiophenes display in response to an applied electric field along the chain direction \cite{10.1103/PhysRevB.81.115401}. 
Finite $\alpha$-helical peptides possess an intrinsic macrodipole arising primarily from the collective orientation of the peptide bonds along the helix axis \cite{Sengupta2005}. The systematic evolution of the HOMO and LUMO energies with increasing helix length is therefore consistent with a Stark-like contribution from the electrostatic potential associated with this intrinsic dipole. As the helix is extended, the increasing macrodipole can generate a larger electrostatic potential difference along the molecular axis, thereby contributing to the observed shifts of the frontier molecular-orbital energies and the narrowing of the H–L gap. The spatial localization of the frontier orbitals and the relationship between this electronic-energy gradient and the helix dipole are examined in the following sections.

\subsection{Spatial Localization and Residue-Resolved Electronic Structure}
\begin{figure}[htbp]
    \centering
    \includegraphics[width=0.8\textwidth]{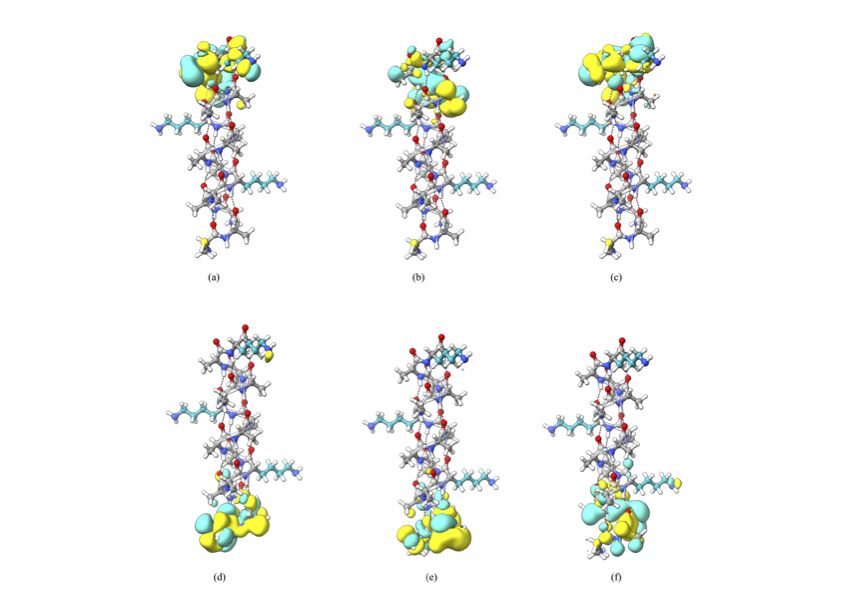}
    \caption{Spatial distribution of the three highest occupied molecular orbitals (a)HOMO, (b)HOMO-1, (c)HOMO-2 and the three lowest unoccupied molecular orbitals (d)LUMO, (e)LUMO+1, and (f)LUMO+2 in C[AAAAK]$_{3}$.}
    \label{fig:3}
\end{figure}

The spatial distributions of the frontier molecular orbitals provide further insight into the electronic-energy variation along the $\alpha$-helical structure. Previous studies of finite $\alpha$-helical peptides have shown that frontier molecular orbitals can preferentially localize toward opposite ends of the helix, a behavior associated with the intrinsic electrostatic polarity of the $\alpha$-helical structure \cite{10.1021/jp302232p}. Figure~\ref{fig:3} shows this localization for C[AAAAK]$_3$. In this polypeptide, the HOMO, HOMO-1, and HOMO-2 are predominantly localized toward the C-terminal region, whereas the LUMO, LUMO+1, and LUMO+2 are localized majorly toward the N-terminal region, where the N-terminal cysteine provides a pronounced contribution to the unoccupied frontier state. 

\begin{figure}[htbp]
    \centering
    \includegraphics[width=0.6\textwidth]{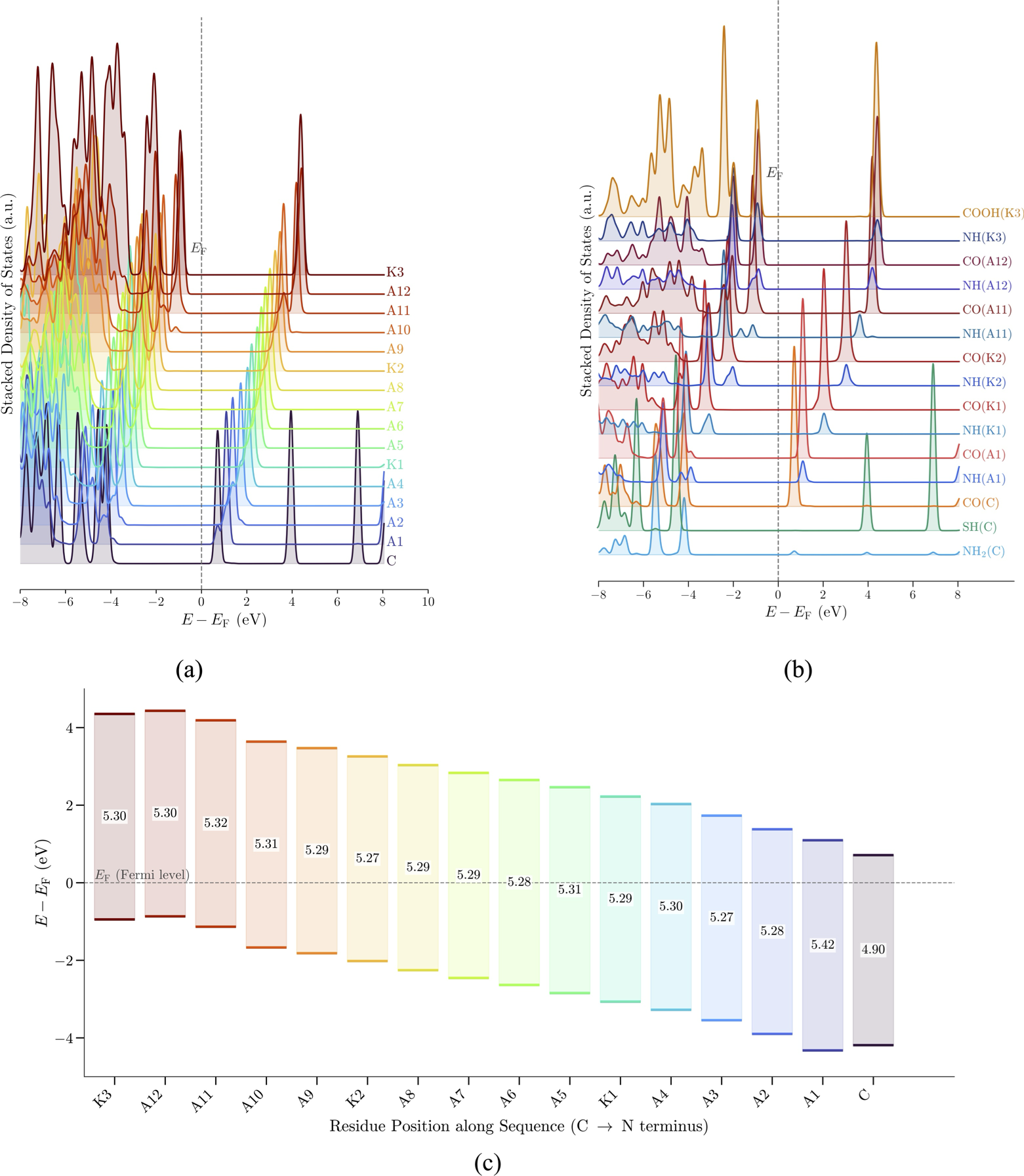}
    \caption{Projected density of states (PDOS) on (a) individual amino acid residues and (b) key functional groups contributing to frontier orbitals in C[AAAAK]$_3$. (c) residue-projected occupied–unoccupied separations calculated for each amino acid position along the sequence from C- to N-terminus.}
    \label{fig:4}
\end{figure}

To provide deeper insight into the contribution of individual amino acids to the electronic structure, we calculated the PDOS for C[AAAAK]$_{3}$. This system is particularly informative because, in addition to the alanine-rich helical backbone, it contains lysine side chains and an N-terminal cysteine. The presence of these chemically distinct groups allows their contributions to the frontier electronic states to be distinguished from those of the peptide backbone. As discussed below using the PDOS, the detailed orbital character is therefore determined by both the position of the residues along the intrinsic electrostatic gradient and their local chemical composition. Figure~\ref{fig:4}a shows the projected states on the individual amino acids , with the labeling starting from the  cysteine (C) unit at the N-terminus and ending with the lysine unit (K3) at the C-terminus. As seen in the plot, cysteine contributes most strongly to the LUMO, while A12 provides the dominant contribution to the HOMO. Moreover, the peaks appearing above the LUMO largely follow the sequence of the amino acids along the helix. Figure~\ref{fig:4}b decomposes the MO state contributions into the selected functional groups of the residues that dominate the relevant orbitals perspective. Specifically, K3, A11, and A12 correspond to HOMO--2 to HOMO, while A1 and C correspond to LUMO+1 and LUMO. The projections were performed onto the CO (COOH for K3), NH (NH$_{2}$ for C), and SH (for C) groups. As seen in the figure, the CO groups of the amino acids provide the dominant contributions to the frontier orbitals, highlighting the central role of the carbonyl units in the electronic structure of the helix. The NH groups generally exhibit smaller contributions, with their projected states showing substantial energetic overlap with those of the carbonyl groups. Interestingly, the pattern of this overlap varies along the helix: near the N-terminus, the NH and CO contributions predominantly coincide within the same residue, whereas toward the C-terminus, the NH contributions increasingly overlap in energy with those of the carbonyl groups of adjacent residues.

The residue-resolved occupied and unoccupied state onsets provide an additional perspective on the local energetic landscape. Figure~\ref{fig:4}c shows the corresponding residue-projected H--L separations along C[AAAAK]$_{3}$. Despite the systematic shifts of the occupied and unoccupied state onsets along the helix, their energy separations remain nearly constant at approximately 5.30 eV for most residues. The largest deviations occur near the N-terminus, where values of 4.90 eV for C and 5.42 eV for A1 are obtained. In contrast, the three residues closest to the C-terminus exhibit nearly identical separations. The deviations from the otherwise regular staircase-like progression of the energy-level onsets near the termini may be attributed to local effects associated with terminal groups and side-chain modifications.

The relatively uniform local gaps indicate that the basic electronic structure of the individual peptide units is preserved along much of the helix, despite the pronounced variation in the absolute energies of the residue-resolved states. Importantly, however, the energetic separation between occupied and unoccupied states associated with different residues can be smaller than the local gap of an individual residue. For example, the energy difference between the occupied state associated with A1 and the unoccupied state associated with A4 is approximately 4.3 eV.

The helical geometry and its hydrogen-bonding network bring residues that are separated along the primary sequence into close spatial proximity. The combination of such spatial proximity with the observed sequence-dependent energetic offsets therefore provides possible lower-energy inter-residue channels compared with the approximately 5 eV local gaps. This electronic arrangement is qualitatively compatible with a stepwise picture of charge transfer through different localized states along the helix.

The approximately 5 eV scale of the residue-projected occupied–unoccupied separations is qualitatively comparable to vertical electronic excitation energies reported from first-principles calculations for isolated amino acids, including an average value of approximately 5.3 eV for alanine\cite{Maul2007}. Moreover, the existence of lower inter-residue energy separations is qualitatively consistent with experimentally reported sub-gap features in transport through polyalanine systems \cite{ha_nanojunctions_2024}. 
While long-range electron transfer in natural conductive biomacromolecules---such as microbial pili and redox enzymes---typically relies on specific redox-active residues (e.g., aromatic $\pi$-stacking or Trp/Tyr hopping) rather than the bare peptide backbone, the underlying $\alpha$-helical scaffold plays a critical, enabling role in these biological systems \cite{10.1021/jp302232p, doi:10.1073/pnas.1512704112}. The intrinsic macrodipole and the resulting spatially varying electrostatic gradient observed in our model helices provide the fundamental directional driving force that may contribute to electrostatic tuning of redox energetics and directional transfer in natural proteins.
The hierarchical electronic structure we describe---characterized by large intra-residue gaps but reduced, hydrogen-bond-mediated inter-residue barriers---creates an electrostatic `staircase' that mirrors the structural and energetic strategies utilized by natural helical proteins to facilitate directional transport. This backbone-driven electrostatic landscape acts as the fundamental template upon which sequence-specific charge-transfer pathways are built, aligning with recent experimental demonstrations where intrinsic dipole fields in biomimetic helices directly accelerate directional proton-coupled electron transfer\cite{10.1021/jacs.6c06120}.

\vspace{-1.0ex}
\subsection{Intrinsic Helix Dipole and Its Relation to the Electronic Energy Gradient}
The systematic spatial differentiation of the MO states can be related to one of the characteristic electrostatic properties of an $\alpha$-helix: its macrodipole. As indicated in Figure\ref{fig:3}, the dipole moment of C[AAAAK]$_{3}$ is oriented approximately along the helix axis from the C-terminal toward the N-terminal region. The same overall orientation is found for the other two peptide families.  

\begin{figure}[h]
    \centering
    \includegraphics[width=0.8\textwidth]{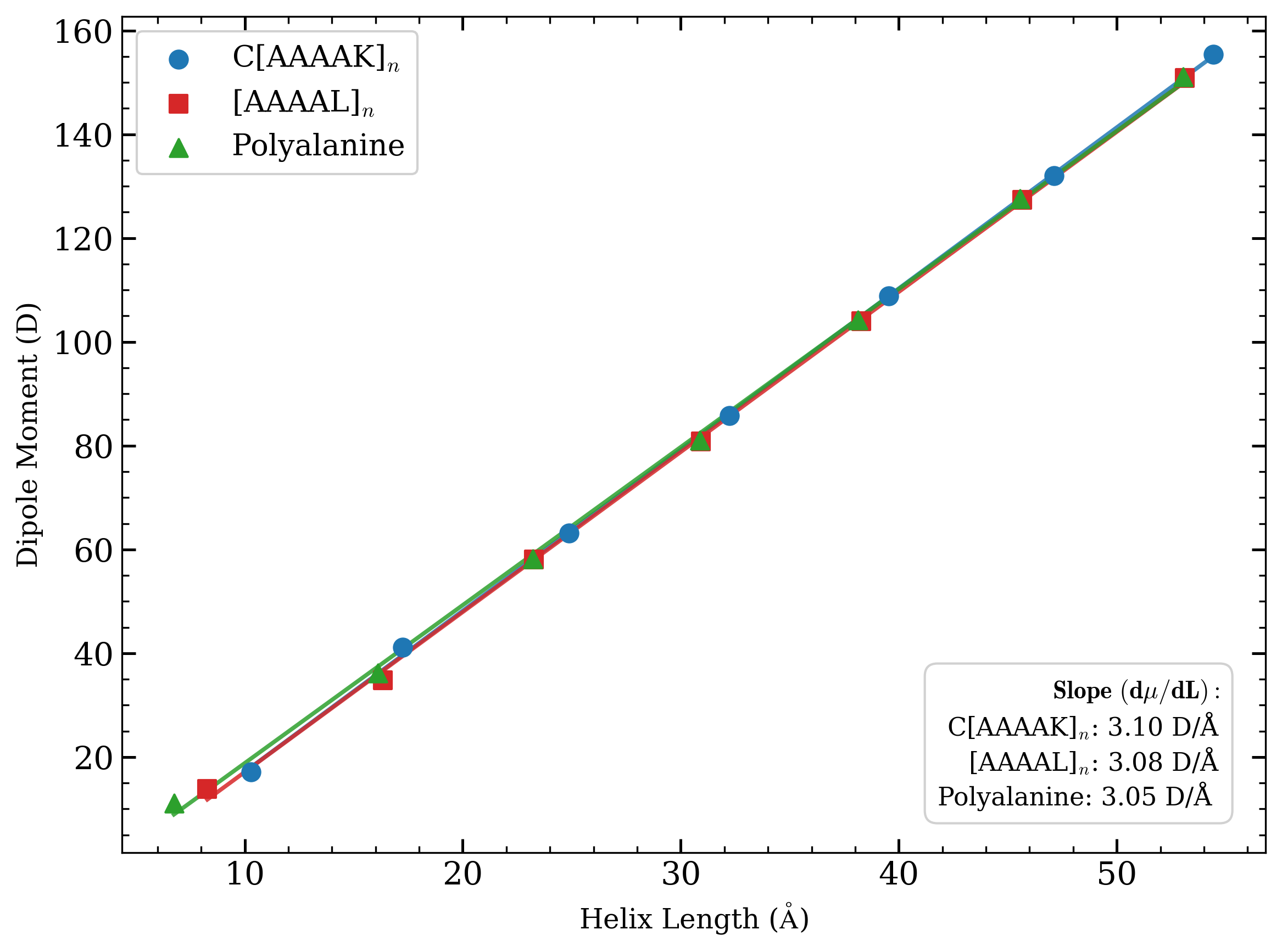}
    \caption{Dipole Moments of the three helices as a function of molecular length.}
    \label{fig:5}
\end{figure}

Figure~\ref{fig:5} presents the magnitude of the molecular dipole moment as a function of chain length. For all three sequence families, the molecular dipole increases approximately linearly with helix length, with comparable slopes $\frac{d\mu}{dL}$, indicating a similar rate of macrodipole buildup across the different sequences. The persistence of this trend across different side-chain chemistries demonstrates that the macroscopic dipole moment scales predictably with chain length, a behavior analogous to the scaling laws observed in other classes of dipolar polymers\cite{HELFER20054361}. This establishes the $\alpha$-helical backbone as a robust and designable structural element for controlling electrostatic properties in biopolymer-based devices.

The linear dependence of the total molecular dipole moment and the length of the model may arise from two distinct effects.
On the one hand, the asymmetric termination at C- and N-terminus might introduce a charge transfer from one end to the other. By increasing the chain length, these charges would be more and more separated and cause an increased dipole moment.
On the other hand, as in the previous section,  where C=O groups of the peptide backbone turned out to be the major contributors to the MO energy levels, the local dipole associated with each C=O bond might be expected to dominate the total dipole moment of the helix.
To test these hypotheses, we determined the partial charges of each amino acid along the peptide as well as the projection of the dipole moment of each C=O group onto the helix axis for C[AAAAK]$_{3}$. The latter was obtained   by scaling the dipole moment of a single C=O bond according to the projection of the bond vector onto the helix axis. The calculation can be expressed as:  
\begin{align}
d_{\text{C=O}} &\approx 1.23 \, \text{\AA}, \\
\mu_{\text{proj}} &= \frac{|\mu|}{d} \, d_{\text{proj}} ,
\end{align}
where $d_{\text{C=O}}$ is the bond length, $d_{\text{proj}}$ is its projection along the helix axis, and $\mu_{\text{proj}}$ is the corresponding dipole projection.  

As summarized in Table~\ref{tab:CO_projection}, 
the partial charges in the middle of the helix nearly vanish with charges below $10^{-2}$ in units of the elementary charge, while the termination groups present charges of 0.05 with the negative partial charge at the N-terminus and the positive one at the C-terminus.
This charge separation generates a dipole moment of about $5.3\,$D which is not only one order of magnitude smaller than the total molecular dipole moment obtained, but also points in the opposite direction. This disproofs the hypothesis, the the dipole moment is caused primarily by the separation of the partially charged terminal groups. 
 
We also notice, that the total charge of the alanine untis Ala2, Ala10 and Ala11 present the largest deviation from zero. This is a consequence of the hydrogen bond from with the terminal amioacid group and their missing linkage to a regular continuing $\alpha$-Helix.  

As furthermore presented in Table~\ref{tab:CO_projection}, all C=O projections have the same sign along the helix axis, reflecting the collective orientation of the backbone carbonyl groups in the $\alpha$-helical structure. Summing these projected contributions yields 62.6 D, which compares quite well with the total molecular dipole moment of 63.1 D obtained for C[AAAAK]$_{3}$. 
Thus, within this bond-projection estimate, the aligned backbone carbonyl groups account for the major fraction of the total molecular dipole.

This comparison should be interpreted as a physically motivated estimate rather than as a unique quantum-mechanical decomposition of the molecular dipole into individual bond contributions. Nevertheless, together with the similar length dependence observed for all three sequence families, it provides strong support for the dominant role of the aligned peptide backbone in establishing the $\alpha$-helix macrodipole.

The increasing molecular dipole with chain length and the opposite-end localization of the frontier orbitals provide a consistent interpretation of the trends observed in Figure\ref{fig:2_all}. As the helix is extended, the collective backbone dipoles generate an increasingly pronounced electrostatic potential difference along the molecular axis. The associated internal field provides a Stark-like contribution to the systematic shift of the MO energies and therefore to the observed reduction of the H–L gap with increasing chain length.

\begin{table}[h!]
\centering
\begin{tabular}{lccccc}
\hline
\textbf{Amino Acid} &
\begin{tabular}{@{}c@{}}Total Residue\\Charge (e)\end{tabular} &
\begin{tabular}{@{}c@{}}Total CO Dipole\\Moment (D)\end{tabular} &
\begin{tabular}{@{}c@{}}CO Dipole Projection\\on Helix Axis (D)\end{tabular} &
\begin{tabular}{@{}c@{}}CO Bond\\Length (\AA)\end{tabular} &
\begin{tabular}{@{}c@{}}CO Projection\\on Helix Axis (\AA)\end{tabular} \\
\hline
Cysteine   & -0.04 & 4.16 & 3.93 & 1.2403 & 1.1717 \\
Alanine 1  &  0.00 & 4.14 & 3.97 & 1.2339 & 1.1842 \\
Alanine 2  & -0.01 & 4.14 & 3.87 & 1.2347 & 1.1544 \\
Alanine 3  &  0.00 & 4.15 & 3.98 & 1.2373 & 1.1864 \\
Alanine 4  &  0.00 & 4.15 & 3.93 & 1.2369 & 1.1721 \\
Lysine 1   &  0.00 & 4.15 & 3.98 & 1.2391 & 1.1878 \\
Alanine 5  &  0.00 & 4.15 & 3.99 & 1.2388 & 1.1891 \\
Alanine 6  &  0.00 & 4.15 & 3.97 & 1.2392 & 1.1842 \\
Alanine 7  &  0.00 & 4.15 & 3.94 & 1.2393 & 1.1738 \\
Alanine 8  &  0.00 & 4.15 & 3.93 & 1.2389 & 1.1727 \\
Lysine 2   &  0.00 & 4.15 & 3.98 & 1.2384 & 1.1859 \\
Alanine 9  &  0.00 & 4.15 & 3.98 & 1.2366 & 1.1866 \\
Alanine 10 &  0.02 & 4.16 & 3.98 & 1.2410 & 1.1856 \\
Alanine 11 &  0.02 & 4.12 & 3.79 & 1.2298 & 1.1306 \\
Alanine 12 &  0.00 & 4.12 & 3.59 & 1.2300 & 1.0710 \\
Lysine 3   &  0.05 & 4.08 & 3.77 & 1.2164 & 1.1235 \\
\hline
\end{tabular}
\caption{Residue charges, C=O bond lengths, dipole moments, and their 
projections on the helix axis for C[AAAAK]$_{3}$.}
\label{tab:CO_projection}
\end{table}
\subsection{Response to an External Electrostatic Field}
\begin{figure}[htbp]
    \centering

    \begin{subfigure}[c]{0.32\textwidth}
        \centering
        \includegraphics[height=6cm,keepaspectratio]{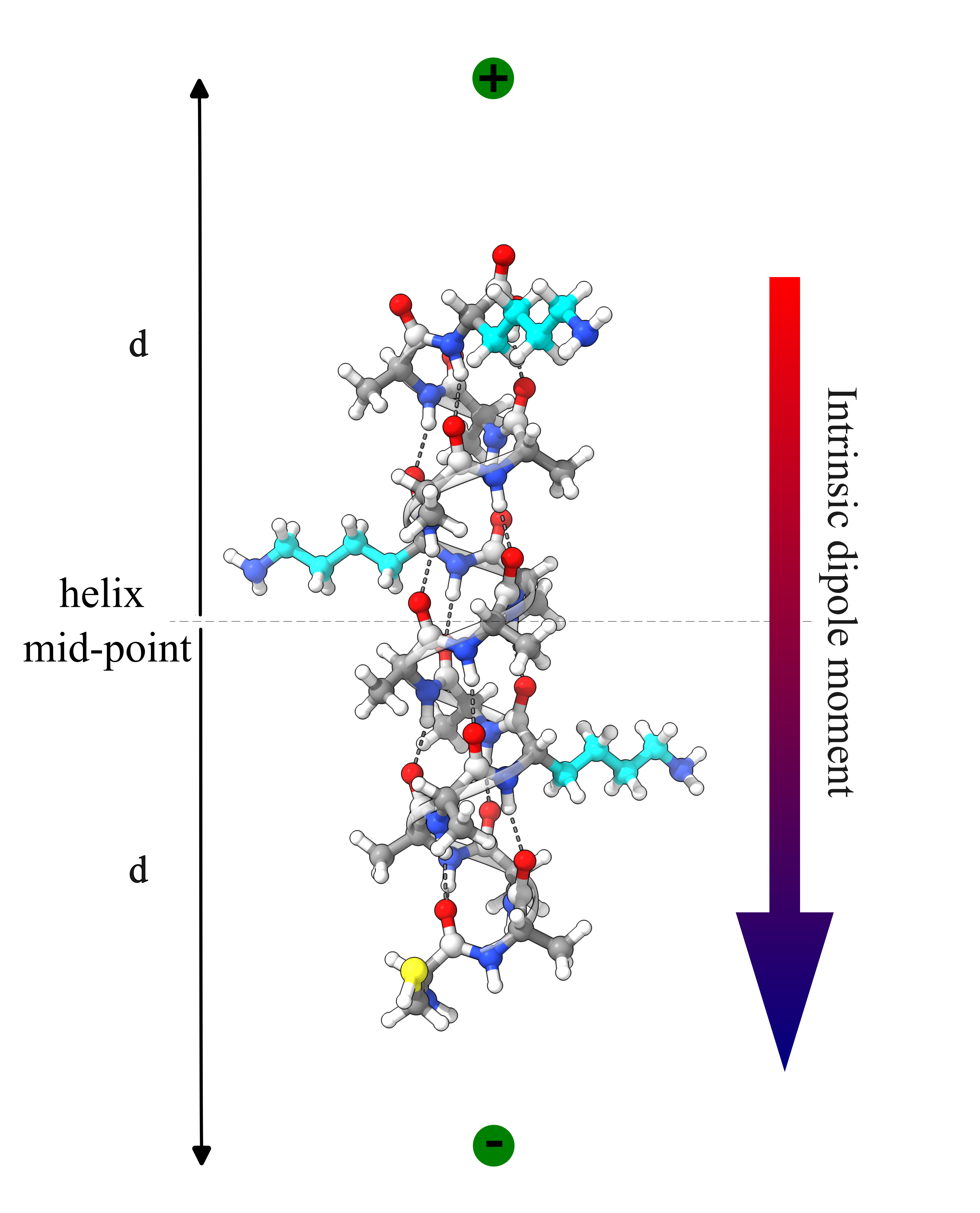}
        \caption{}
        \label{fig:6a}
    \end{subfigure}
    \hfill
    \begin{subfigure}[c]{0.65\textwidth}
        \centering
        \includegraphics[height=6cm,keepaspectratio]{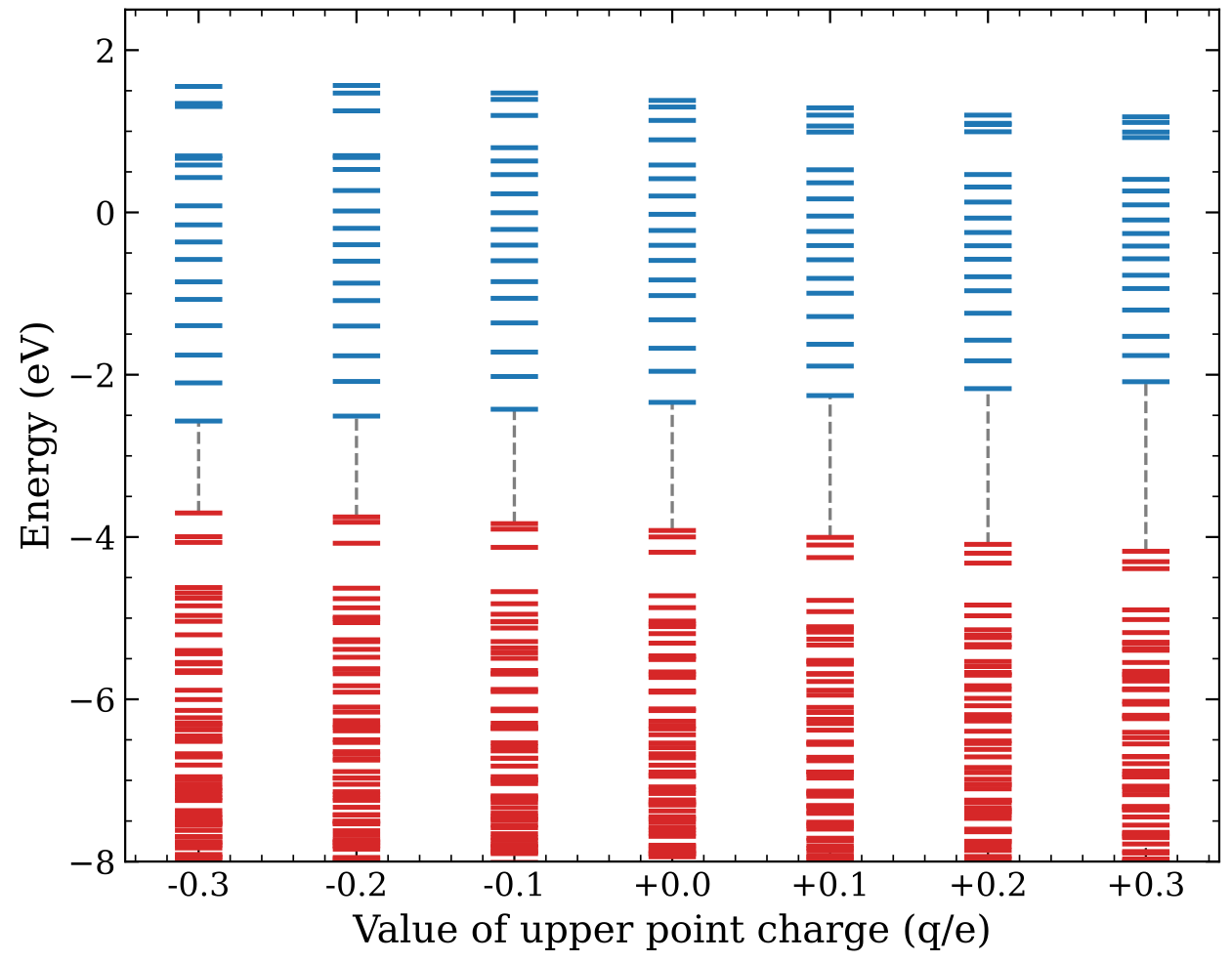}
        \caption{}
        \label{fig:6b}
    \end{subfigure}

    \caption{(a) Point charges at top and bottom of the molecule. d is the distance from the charge to the center of the helix. (b) the energy spectrum for different charge magnitudes on C[AAAAK]$_3$.}
    \label{fig:6}
\end{figure}

 In previous studies we showed that the measurement conditions to determine the conductance of a single molecule in a mechanically controlled break junction (MCBJ) \cite{Slawig2020} and similar conductance measurements in a scanning tunneling microscope (STM) \cite{Nguyen2024} can be reproduced well by suitable simulation approaches which resolve the slow structural relaxations in the MCBJ \cite{Lokamani2023}, whereas for the fast STM the applied electrostatic perturbation by the tip apex is the relevant factor \cite{Kibsgaard2008}. 
Thus, to evaluate the influence of the dipole moments on the STM measurements of $\alpha$-helix polypeptides \cite{Nguyen2024} we investigated the electronic response of the polypeptide C[AAAAK]$_{3}$ to an external electrostatic perturbation. 
For this purpose we placed two point charges of opposite sign and with varying magnitudes (fractions of the elementary charge), along the helix axis at both ends of the C[AAAAK]$_{3}$ polypeptide. The configuration of the point charges is illustrated in Figure~\ref{fig:6a}, where the upper part of the helix corresponds to the C-terminus. This construction provides a controllably non-uniform electrostatic field along the molecular axis and serves as a simplified representation of the electrostatic perturbation that can occur in a molecular-junction or STM-like environment.

 We first considered point charges of different magnitudes positioned 20~\AA{} away from the molecular center at the C-terminus, with an oppositely signed charge placed symmetrically at the N-terminus. This choice poses the point charges at a distance of about 8~\AA{} away from the termini of the polypeptide. Assuming a typical bond distance between gold and lighter elements of about 2 ~\AA{}, this model represents the field of a hemi-spherical gold tip with a diameter  of 12~\AA{}, equivalent to five gold atoms, by the point charge in the center of the apex.
As a reference, two elementary charges of magnitude 1 q/e, opposite sign, and separated by 40~\AA{} in vacuum would correspond to a dipole moment of approximately 192.1 D. This amounts to about three times the intrinsic dipole moment of C[AAAAK]$_{3}$, thus, we varied the point charges in a range of $\pm$0.3 q/e.

Figure~\ref{fig:6b} shows the energy spectrum around the HOMO and LUMO levels for charge magnitudes between -0.3 and +0.3 q/e in steps of 0.1 q/e. Changing the magnitude and polarity of the external charges systematically shifts the occupied and unoccupied frontier levels and consequently modifies the H–L gap. Table~\ref{tab:point_charges} shows that the H–L gap changes from 1.13 eV at one field polarity to 2.09 eV after reversal of the charge polarity, compared with 1.58 eV in the absence of the external charges.

\begin{table}[h!]
\centering
\begin{tabular}{|c|c|c|c|}
\hline
\textbf{Point Charges} & \textbf{Fermi Level (eV)} & \textbf{HOMO-LUMO Gap (eV)} & \textbf{Dipole moment (Debye)} \\
\hline
-0.3 & -3.13 & 1.13 & 59.62 \\
-0.2 & -3.13 & 1.24 & 61.83 \\
-0.1 & -3.13 & 1.41 & 62.52 \\
0.0  & -3.13 & 1.58 & 63.16 \\
0.1  & -3.13 & 1.75 & 63.76 \\
0.2  & -3.13 & 1.92 & 64.34 \\
0.3  & -3.13 & 2.09 & 64.93 \\
\hline
\end{tabular}
\caption{Variation of Fermi level, H-L gap, and dipole moment with point charges.}
\label{tab:point_charges}
\end{table}

The response depends on the relative orientation of the imposed electrostatic field and the intrinsic molecular dipole. For the orientation parallel to the intrinsic dipole, increasing the external perturbation reduces the H–L gap and is accompanied by a decrease of the molecular dipole from 63.1 D to 59.6 D at the largest investigated charge magnitude. Reversing the field orientation increases the H–L gap and produces a modest increase of the molecular dipole, reaching approximately 64.9 D. Thus, the intrinsic and externally imposed electrostatic potentials interact directly in determining the frontier-state separation.

 These findings are consistent with the level diagrams obtained for the infinitely periodic structures (see SI). In the absence of terminating groups their contribution to a dipole moment vanishes. Furthermore, a finite macroscopic dipole moment is not compatible with the applied periodic boundary conditions, thus an internal compensation of the local dipole moments of the C=O groups is achieved and no net intrinsic electrostatic field is built up. In this model-immanent absence of the intrinsic dipole moment, the relative shifts of the electronic levels along the helix also vanish and the periodic structure has a much increased HOMO-LUMO splitting comparable with the HOMO-LUMO splitting of a single amino acid.

The systematic change of the HOMO and LUMO levels of the finite polypeptide with the sign of the applied electrostatic potential should lead to some asymmetry in the experimental current-voltage curves. In particular, the onset of the current should occur at different absolute values for positive and negative applied voltages, thus the helix could act as a molecular rectifier within a limited bias window. A closer inspection of the measured current-voltage curves in ref.~\cite{Nguyen2024} indicates indeed the occurrence of such asymmetries with respect to the inflection point at zero applied bias, but the magnitude may still be within the error bars of the measurements.

The changes of the molecular dipole are substantially smaller than the dipole associated with the external point-charge configuration itself. The molecular dipole also retains its original orientation throughout the investigated field range. These results indicate that the intrinsic helix polarity remains a robust component of the electronic structure under the applied perturbations considered here, although its magnitude and the frontier-level energies can be modulated by the external field.

 As the local bonding geometry proved particularly relevant for the comparison with experimental results from long measurement times we next kept the point charges of at a magnitude of 0.3 q/e and varied their symmetric distance to the helix from 18 to 23~\AA{} relative to the molecular center. The resulting Fermi energies, H–L gaps, and dipole moments are summarized in Table~\ref{tab:distance}. The DFTB Fermi energy remains similar for all the distances. In contrast, the H–L gap continues to vary systematically over the entire investigated distance range, decreasing from 2.26 eV at 18~\AA{} to 1.93 eV at 23~\AA{}. The molecular dipole also changes gradually, from 65.6 to 64.3 D.

\begin{table}[h!]
\centering
\begin{tabular}{|c|c|c|c|}
\hline
\textbf{Distance} & \textbf{Fermi} & \textbf{HOMO--LUMO} & \textbf{Dipole} \\
\textbf{from center (\AA)} & \textbf{level (eV)} & \textbf{gap (eV)} & \textbf{moment (Debye)} \\
\hline
18 & -3.13 & 2.26 & 65.6  \\
19 & -3.13 & 2.16 & 65.2 \\
20 & -3.13 & 2.09 & 64.9 \\
21 & -3.13 & 2.03 & 64.7 \\
22 & -3.13 & 1.97 & 64.5 \\
23 & -3.13 & 1.93 & 64.3 \\
\hline
\end{tabular}
\caption{Variation of dipole moment, Fermi level, and HOMO--LUMO gap with distance of point charges from the molecular center.}
\label{tab:distance}
\end{table}

A distance of 20 Å was subsequently used as a practical compromise that produces a clear electronic perturbation while limiting the associated structural deformation. The applied electrostatic field also modifies the molecular geometry. Comparison of the C=O bond projections before and after field application shows small angular changes whose magnitude generally increases toward the C-terminal region (see SI).
For the selected charge configuration, the estimated field magnitude is spatially nonuniform, with values of approximately $6.66 \times 10^{8}$~V/m near the C-terminus,  $1.98 \times 10^{8}$~V/m near the molecular center, and $3.35 \times 10^{8}$~V/m near the N-terminus. Consequently, the larger geometrical response observed toward the C-terminal region cannot be attributed solely to greater intrinsic flexibility of those residues, because this region also experiences the strongest and most inhomogeneous part of the local electric field. The observed deformation therefore reflects the combined effects of the spatially varying field and the local structural response of the peptide.

Overall, the external-field calculations demonstrate that the frontier electronic structure of the $\alpha$-helix is tunable by an electrostatic perturbation while the intrinsic molecular polarity remains present over the investigated range.

\section{Conclusion}

In this work, we employed density-functional tight-binding (DFTB+) simulations to elucidate how side-chain chemistry and helix length govern the electronic structure of designed $\alpha$-helical polypeptides, utilizing three model systems: polyalanine [AAAAA]$_n$, asymmetric leucine-modified sequences [AAAAL]$_n$, and N-terminally cysteine-capped, lysine-terminated variants C[AAAAK]$_n$. Our calculations reveal that molecular orbital energies evolve in an approximately semi-linear fashion with increasing residue number---a signature of an intrinsic macrodipole field arising from the collective alignment of backbone amide groups. This internal field produces a Stark-like splitting of frontier orbitals that persists even under externally applied electric fields, demonstrating the robustness of the helix dipole as a defining electronic feature of these biomacromolecules.

Critically, we find that realistic external fields---such as those generated by an STM tip or at a bio-electronic interface---induce structural and electronic perturbations that are spatially non-uniform along the helix. Electrostatic forces concentrate near the termini,where they act synergistically with the greater intrinsic flexibility of the C=O bond angles, making field-induced distortions most pronounced in these regions, while the helical core remains comparatively rigid. This terminus-localized response mirrors the spatial profile of experimentally relevant tip--sample fields and provides a molecular-scale rationale for observed STM conductance asymmetries in peptide junctions, highlighting the structural robustness of the $\alpha$-helical scaffold for bio-hybrid device integration.

Furthermore, our analysis of the local density of states reveals a staircase-like organization of localized electronic states along the helical axis that may provide an energetic framework for stepwise charge transfer. While individual peptide units exhibit large local H--L gaps ($\sim$5~eV), the hydrogen-bond network bridging adjacent residues across the helical pitch creates intermediate states that effectively reduce the transport barrier between units. This hierarchical electronic structure---featuring large intra-residue gaps and reduced barriers between hydrogen-bonded residues---creates an electronic `staircase' that facilitates long-range charge transport. This mechanism not only accounts for sub-gap optical excitations observed in illuminated peptides but also mirrors the long-range electron transfer strategies utilized by natural conductive biomacromolecules, such as microbial pili and redox proteins.

Together, these findings establish a hierarchy of structure–property relationships in $\alpha$-helical polypeptides, in which side-chain and terminal modifications perturb specific frontier states while the ordered peptide backbone generates a chain-length-dependent macrodipole that organizes the global electronic-energy landscape. The coexistence of large residue-projected occupied–unoccupied separations with smaller inter-residue energetic offsets further reveals a staircase-like electronic organization that may be relevant to stepwise charge-transfer processes, while externally imposed electrostatic perturbations modulate frontier-level separations without eliminating the intrinsic helix polarity. 

From a macromolecular design perspective, separating local chemical effects from collective backbone electrostatics provides a route for tuning dipolar and frontier-electronic properties through sequence and chain length. These structure–property relationships provide a fundamental framework for designing sequence-defined polypeptide materials with controllable electrostatic response and electronic-state organization, with potential relevance to molecular wires, bioelectronic interfaces, field-responsive sensors, and future studies of charge- and spin-selective transport.

The present results provide an electronic-structure basis for viewing $\alpha$-helical peptides as robust yet chemically adaptable building blocks for functional biomolecular materials. While the peptide backbone establishes a persistent dipolar architecture and a characteristic spatial organization of electronic states, sequence variation and external electrostatic fields provide additional degrees of freedom for modulating frontier-state energies and localization. This separation between a robust backbone-defined electronic framework and locally tunable electronic features suggests a route toward sequence-level control of peptide electronic functionality. Future studies should determine how these intrinsic design principles evolve in more realistic environments, where solvation, counter ions, conformational fluctuations, intermolecular interactions, and interfaces with electrodes or other materials may reshape the electrostatic potential and frontier electronic structure. Extending such analyses from isolated helices to peptide assemblies and hybrid interfaces will be particularly important for connecting molecular-scale electronic organization to experimentally measurable charge-transfer, transport, and field-responsive behavior. More broadly, combining computationally efficient quantum-mechanical approaches with systematic sequence variation could enable predictive screening of helical peptides in which structural stability, dipolar strength, and electronic-state localization are tuned independently or cooperatively. Such a framework may ultimately support the rational development of sequence-defined peptide materials for bioelectronic interfaces, molecular sensing, and electronically active biomimetic systems.

\section*{Supplementary Material}
\label{Supplementary Material}
Supplementary material associated with this article contains additional tests for selected polypeptides with D3 correction, as well as additional PDOS plots demonstrating the consistency of amino acids and functional groups across the C[AAAAK]$_n$ series, alongside data tables on the geometrical response of C[AAAAK]$_3$ under an applied electric field.

\section*{Acknowledgements}
\label{Acknowledgements}
The authors gratefully acknowledge the Deutsche Forschungsgemeinschaft for funding via projects B02 and iRTG of the Collaborative Research Center Transregio TRR 386 (HYPMOL) under grant number 51 466 4767 and by the European Regional Development Fund (ERDF) of the Free State of Saxony via the SAX-QT project.

\section*{Data Availability Statement}
\label{Data Availability Statement}
The data that supports the findings of this study are available within the article and its supplementary material.

\bibliographystyle{unsrt}      
\bibliography{./sample}      

\newpage
\renewcommand{\thesection}{S\arabic{section}}
\renewcommand{\thefigure}{S\arabic{figure}}
\renewcommand{\thetable}{S\arabic{table}}

\setcounter{figure}{0}
\setcounter{table}{0}
\setcounter{section}{0}

\section*{Supporting Information}

\section{Properties of the Infinite Periodic Polyalanine Chain}

\begin{figure}[bthp]
    \centering
    \includegraphics[width=0.7\textwidth]{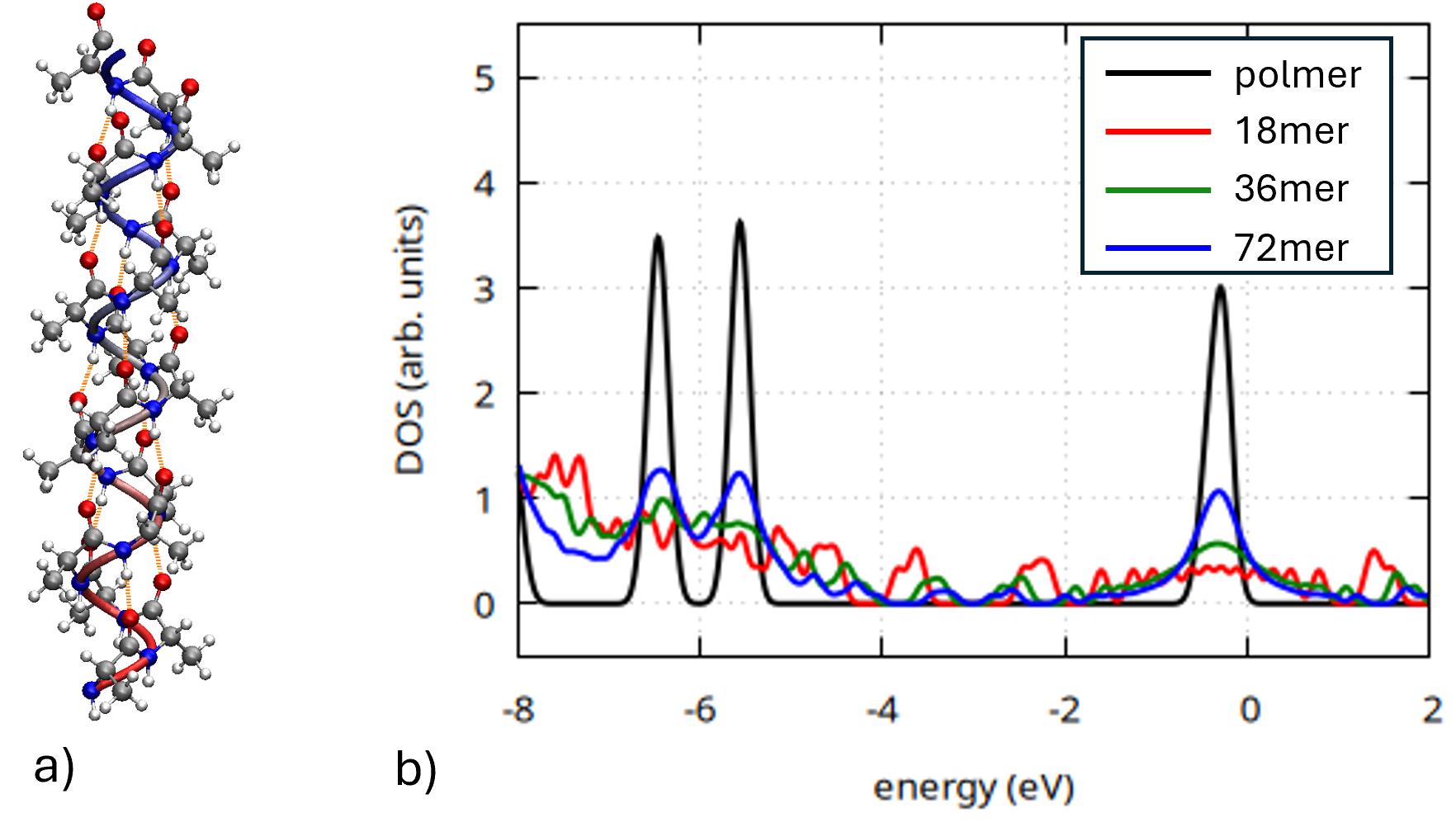}.png
    \caption{Electronic structure of the infinite 1D periodic polyalanine helix compared to finite models. 
    Structural representation of the infinite $\alpha$-helical unit cell comprising 18 alanine residues over 5 helical turns subject to 1D periodic boundary conditions (a). 
    Density of states (DOS) of the infinite polymer compared to finite polyalanine helices of varying chain lengths (18-, 36-, and 72-mer) (b).}
    \label{fig:infinite_polymer}
\end{figure}

To disentangle termination and end-group effects from the intrinsic electronic structure of the $\alpha$-helix backbone, an infinite polyalanine helix subject to 1D periodic boundary conditions along the helical axis was calculated. To correctly represent the geometrical properties of an $\alpha$-helix structure, the unit cell consisted of 18 alanine units spanning 5 helical turns. The atomic structure of the repeating helical unit is presented in Figure~\ref{fig:infinite_polymer}a).

In contrast to the finite models (see Figure 2 in the main manuscript), the electronic structure of the infinite model revealed flat valence and conduction bands with nearly zero dispersion and a bandgap of $5.14\text{ eV}$. Figure~\ref{fig:infinite_polymer}b) displays the density of states (DOS) obtained for the infinite model alongside those of finite models with 18, 36, and 72 alanine units. It can be seen that for increasingly long finite models, the DOS peaks progressively emerge in the energy range corresponding to the bands of the infinite model. 

Importantly, however, the localized end states that narrow the HOMO--LUMO gap in finite models persist across all investigated oligomer lengths. Bulk-like electronic features begin to dominate only for models exceeding approximately 36 units.
Since experimentally realized polypeptides are frequently shorter than this threshold, accounting for termination and end effects as discussed in the main manuscript is essential for investigating their electronic properties.

\section{Sensitivity of dipolar and frontier electronic properties to D3 dispersion}
\label{sec:d3_tests}

The systematic calculations reported in the main manuscript were performed without an explicit dispersion correction. To assess whether this choice affects the central chain-length- and sequence-dependent conclusions, representative D3-corrected calculations were carried out for [AAAAA]$_3$, [AAAAA]$_6$, [AAAAL]$_6$, and C[AAAAK]$_6$. Table~\ref{tab:d3_tests} compares the total dipole moment, Fermi level, and HOMO--LUMO (H--L) gap obtained without D3 and after D3 treatment. For C[AAAAK]$_6$, both a D3 single-point calculation on the geometry optimized without D3 and a full D3 geometry optimization were performed, allowing the effect at fixed geometry to be distinguished from changes induced by structural relaxation.

\begin{table}[htbp]
    \centering
    \small
    \caption{Sensitivity of the dipole moment, Fermi level,
    and HOMO--LUMO gap to dispersion treatment for
    representative peptide helices.}
    \label{tab:d3_tests}

    \setlength{\tabcolsep}{4pt}
    \renewcommand{\arraystretch}{1.15}

    \begin{tabularx}{\linewidth}{
        l
        >{\raggedright\arraybackslash}X
        ccc
    }
        \toprule
        \textbf{System} &
        \textbf{Disp. rel.} &
        \textbf{Dipole (D)} &
        \textbf{$E_F$ (eV)} &
        \textbf{H--L gap (eV)} \\
        \midrule

        C[AAAAK]$_3$ & UFF  & 63.1 & -3.13 & 1.58 \\
        C[AAAAK]$_3$ & no dispersion & 63.6 & -3.13 & 1.58 \\
        C[AAAAK]$_3$ & D3 & 66.5 & -3.07 & 1.19 \\
        
        \addlinespace

        {[AAAAA]$_3$} & UFF & 58.2 & -3.17 & 1.67 \\
        {[AAAAA]$_3$} & no dispersion & 59.0 & -3.13 & 1.81 \\
        {[AAAAA]$_3$} & D3 & 61.5 & -3.11 & 1.22 \\

        \addlinespace

        {[AAAAA]$_6$} & UFF  & 127.5 & -3.14 & 1.03 \\
        {[AAAAA]$_6$} & no dispersion & 132.2 & -3.11 & 1.00 \\
        {[AAAAA]$_6$} & D3  & 137.2 & -3.07 & 0.57 \\

        \addlinespace

        {[A]$_\infty$} &
        {no D3} &
        {---} &
        {-3.138} &
        {5.116} \\

        {[A]$_\infty$} &
        {D3 optimized} &
        {---} &
        {-2.853} &
        {5.126} \\

        \addlinespace

        {[AAAAL]$_3$} & UFF  & 58.0 & -3.15 & 1.65 \\
        {[AAAAL]$_3$} & no dispersion & 58.0 & -3.12 & 1.65 \\
        {[AAAAL]$_3$} & D3  & 61.1 & -3.10 & 1.25 \\

        \bottomrule
    \end{tabularx}
\end{table}
For C[AAAAK]$_6$, inclusion of D3 at the single-point level leaves the calculated quantities essentially unchanged: the dipole moment changes from 137.11 to 136.95~D, $E_F$ from $-3.047$ to $-3.050$~eV, and the H--L gap from 1.194 to 1.203~eV. In contrast, geometry optimization with D3 gives a dipole moment of 141.35~D and reduces the H--L gap to 0.708~eV, while $E_F$ remains close to $-3$~eV. The substantially larger change after D3 relaxation therefore indicates that the quantitative modification of the frontier-level separation is associated predominantly with dispersion-induced structural relaxation rather than with the D3 treatment at fixed geometry.

Importantly, the principal chain-length dependence is retained. Without D3, increasing the polyalanine chain from [AAAAA]$_3$ to [AAAAA]$_6$ increases the dipole moment from 58.78 to 131.96~D and decreases the H--L gap from 1.909 to 1.249~eV. The D3-optimized structures show the same behavior: the dipole moment increases from 62.56 to 138.28~D, while the H--L gap decreases from 1.311 to 0.634~eV. Thus, although D3 relaxation modifies the absolute frontier-level separations, it does not change the qualitative correlation between increasing chain length, increasing molecular dipole, and narrowing of the H--L gap. 
This length dependence vanishes for the infinite polymer ($\text{[A]}_\infty$), where full D3 relaxation leaves the bandgap virtually unaffected ($5.116$ vs. $5.126\text{ eV}$). This demonstrates that dispersion forces play a minor role in the bulk backbone and confirms that the larger D3-induced shifts in finite helices stem from structural relaxations at the unconstrained, non-compensated termini where the frontier orbitals reside.

The D3-optimized six-repeat systems also retain similar overall dipolar and electronic characteristics across the three sequence families. The dipole moments of [AAAAA]$_6$, [AAAAL]$_6$, and C[AAAAK]$_6$ are 138.28, 136.20, and 141.35~D, respectively, while the corresponding H--L gaps are 0.634, 0.743, and 0.708~eV. Their Fermi levels likewise remain within a narrow range ($-3.054$ to $-3.030$~eV). These results are consistent with the main-manuscript conclusion that side-chain and terminal modifications introduce quantitative perturbations while preserving the common backbone-dominated dipolar and frontier-electronic behavior. Overall, the D3 tests therefore support the robustness of the principal chain-length- and sequence-dependent trends discussed in the main text.

\section{Consistency of amino acid and functional groups contributions across the three types of $\alpha$-helix series }
\label{sec:pdos_consistency}

\begin{figure}[htbp]
    \centering

    \begin{subfigure}[b]{0.46\textwidth}
        \centering
        \includegraphics[width=\linewidth]{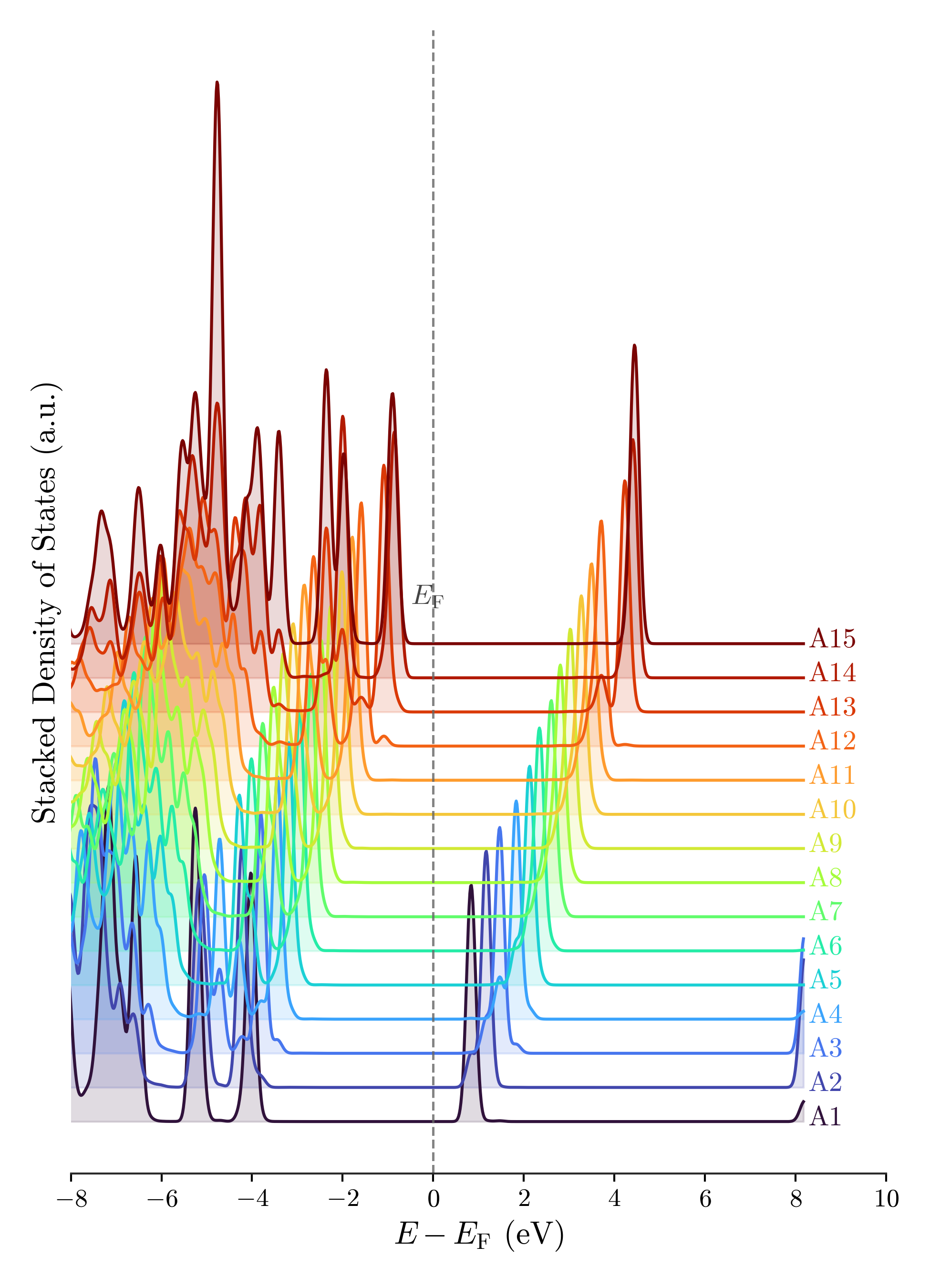}
        \caption{}
        \label{fig:PDOS_PA}
    \end{subfigure}
    \hfill
    \begin{subfigure}[b]{0.46\textwidth}
        \centering
        \includegraphics[width=\linewidth]{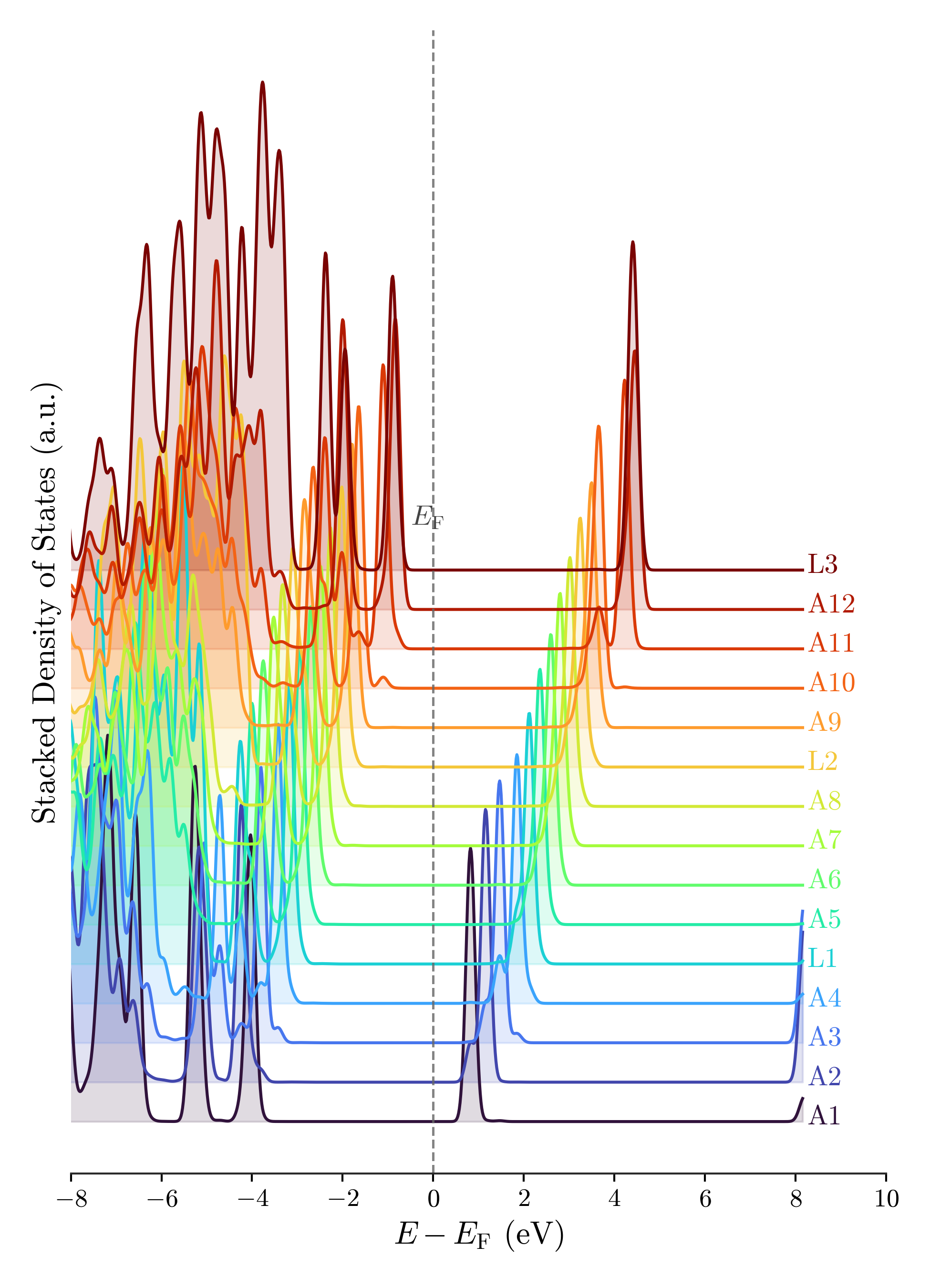}
        \caption{}
        \label{fig:PDOS_AAAAL}
    \end{subfigure}

    \vspace{0.3cm}

    \begin{subfigure}[b]{0.46\textwidth}
        \centering
        \includegraphics[width=\linewidth]{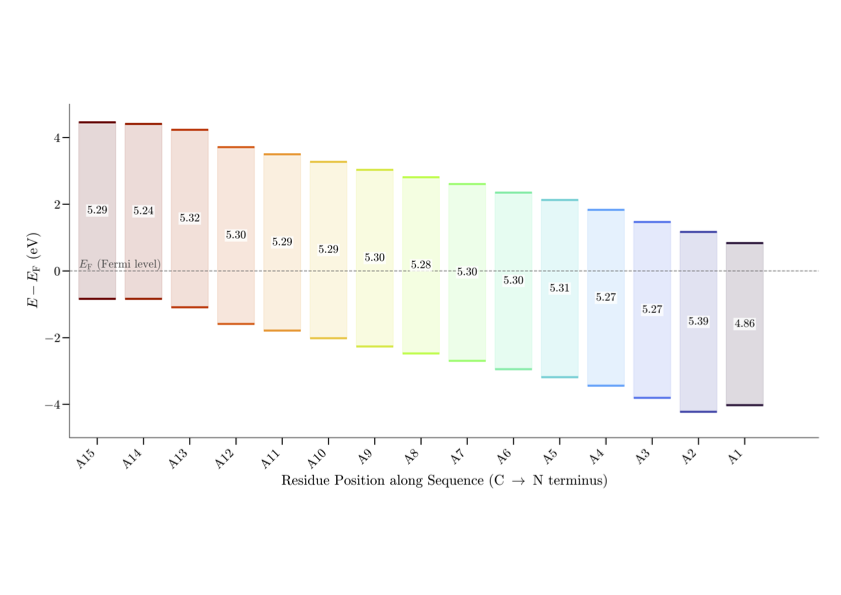}
        \caption{}
        \label{fig:gap_PA}
    \end{subfigure}
    \hfill
    \begin{subfigure}[b]{0.46\textwidth}
        \centering
        \includegraphics[width=\linewidth]{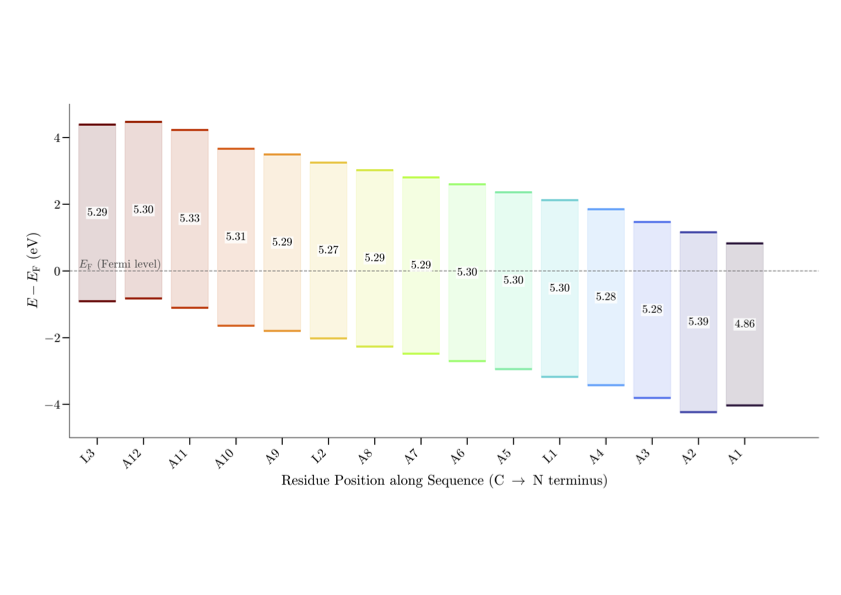}
        \caption{}
        \label{fig:gap_AAAAL}
    \end{subfigure}

    \caption{Residue-resolved electronic structure of
    [AAAAA]$_3$ and [AAAAL]$_3$.
    Projected densities of states (PDOS) for
    (a) [AAAAA]$_3$ and (b) [AAAAL]$_3$.
    Corresponding residue-resolved HOMO--LUMO energy
    separations for (c) [AAAAA]$_3$ and
    (d) [AAAAL]$_3$.
    }

    \label{fig:PDOS_comparison}
\end{figure}

In the main manuscript (Fig.~4), we presented the projected density of states (PDOS) for the C[AAAAK]$_3$ peptide, highlighting the characteristic electronic contributions of the amino acids, the peaks appearing above the LUMO largely follow the sequence of the amino acids along the helix. To demonstrate that this characteristic is consistent in the other two $\alpha$-helices: [AAAAL]$_3$ and [AAAAA]$_3$, we have presented the residue-resolved PDOS and the residue-projected occupied–unoccupied separations calculated for each amino acid position along the sequence from C- to N-terminus in Fig.~\ref{fig:PDOS_comparison}.

\begin{figure}[htbp]
    \centering
    \includegraphics[width=0.95\textwidth]{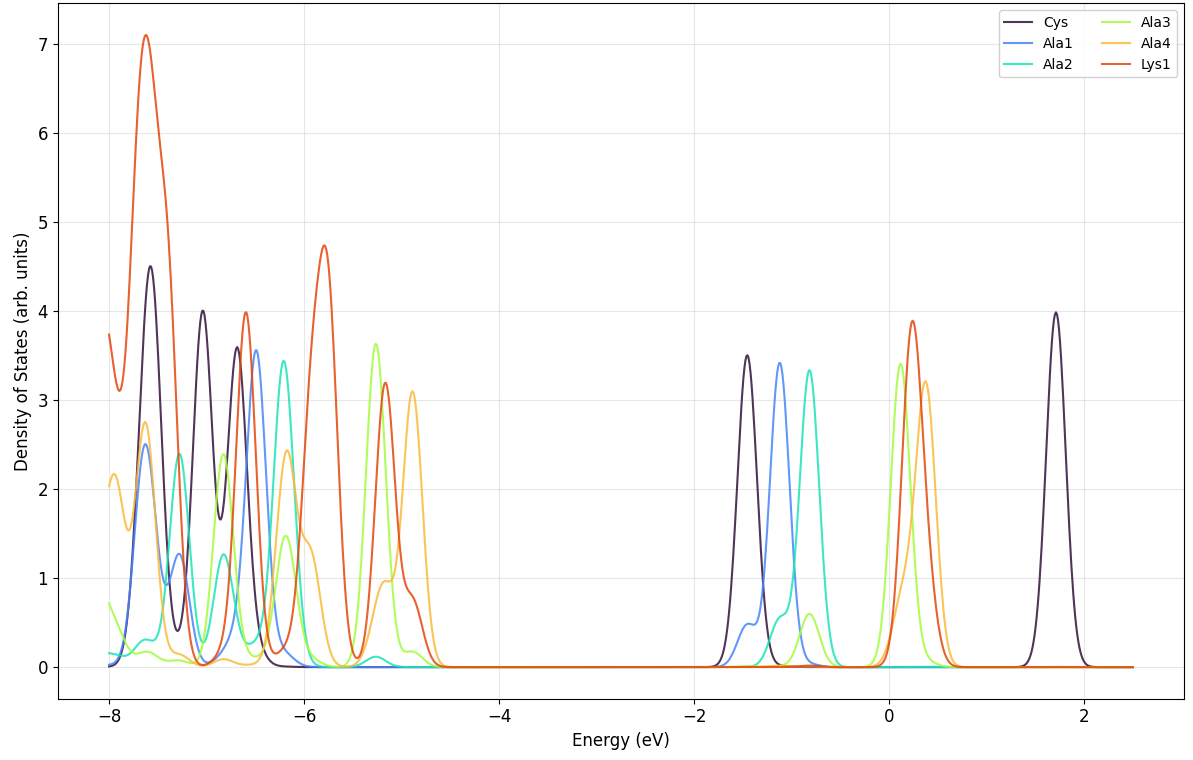}
    \caption{
        PDOS  for C[AAAAK] , decomposed by amino acid type. 
    }
    \label{fig:si_pdos_n1}
\end{figure}

\begin{figure}[htbp]
    \centering
    \includegraphics[width=0.95\textwidth]{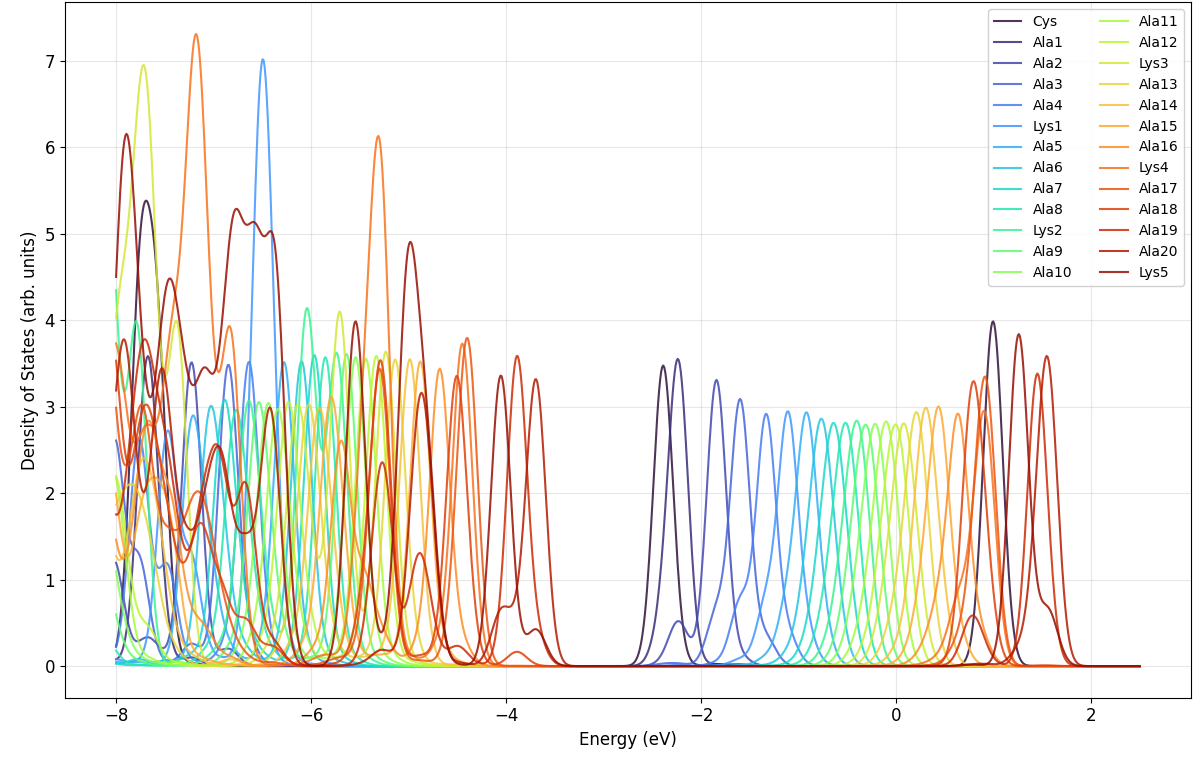}
    \caption{
        PDOS for C[AAAAK]$_5$, decomposed by amino acid type. 
    }
    \label{fig:si_pdos_n5}
\end{figure}

Moreover, in order to show the consistency across the series, we have shown complementary PDOS analyses for two additional representatives: the shortest member C[AAAAK] ($n=1$) and an extended variant C[AAAAK]$_5$ ($n=5$) in Figs.~\ref{fig:si_pdos_n1} and \ref{fig:si_pdos_n5}.


\medskip
\noindent\textbf{Functional group contribution in frontier molecular orbitals}

\begin{figure}[htbp]
    \centering
    \begin{subfigure}[b]{0.485\textwidth}
        \centering
        \includegraphics[width=\textwidth]{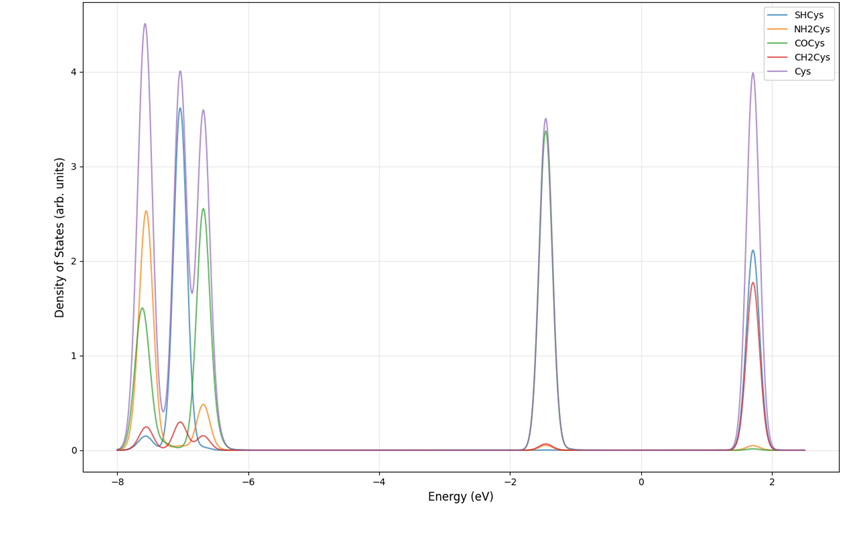}
        \caption{Cysteine functional groups}
        \label{fig:fg_n1_cys}
    \end{subfigure}
    \hfill
    \begin{subfigure}[b]{0.47\textwidth}
        \centering
        \includegraphics[width=\textwidth]{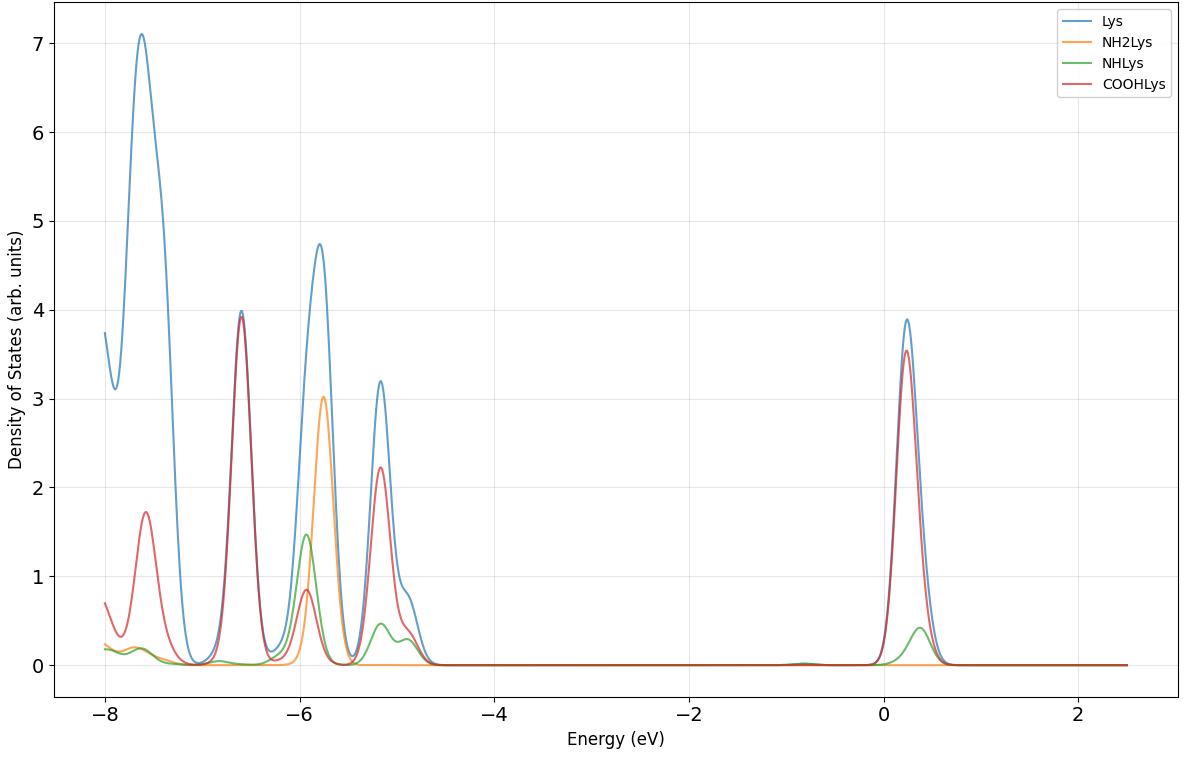}
        \caption{Lysine functional groups}
        \label{fig:fg_n1_lys}
    \end{subfigure}
    \caption{
        Functional group-projected PDOS for C[AAAAK]. 
        \textbf{(a)} Cysteine
        \textbf{(b)} Lysine
    }
    \label{fig:si_fg_n1}
\end{figure}

\begin{figure}[htbp]
    \centering
    \begin{subfigure}[b]{0.51\textwidth}
        \centering
        \includegraphics[width=\textwidth]{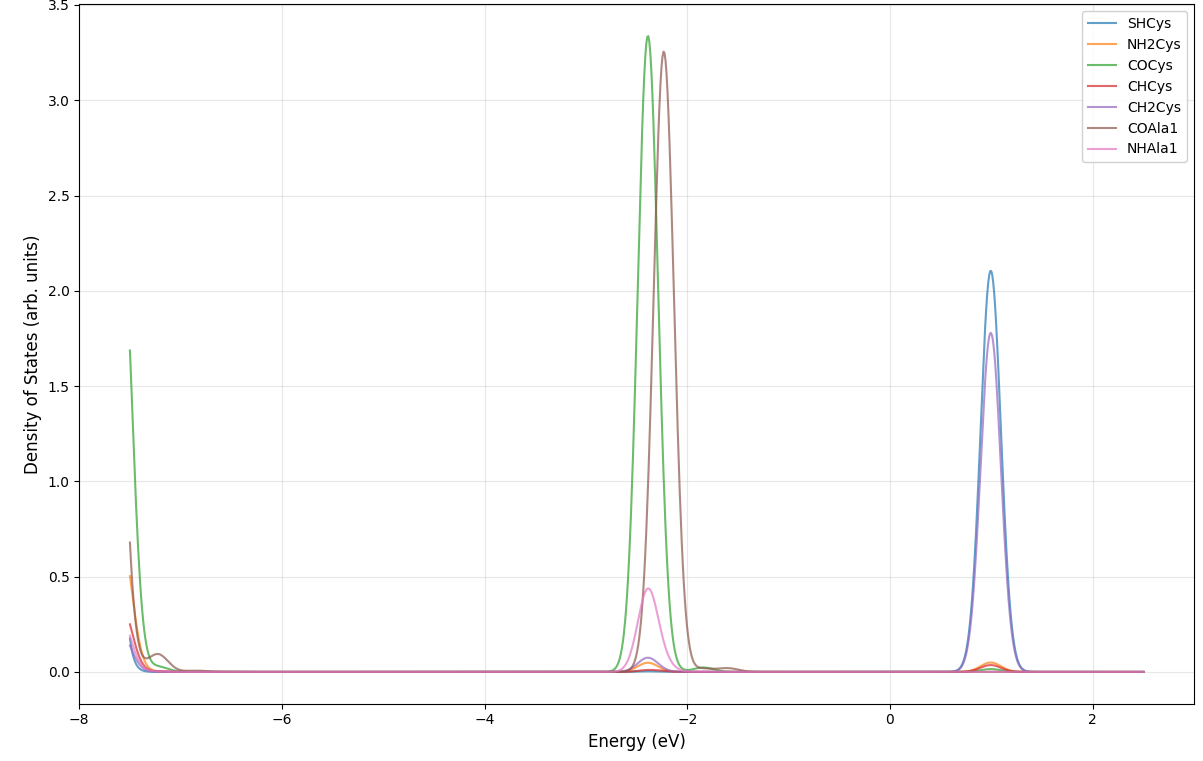}
        \caption{Cysteine functional groups}
        \label{fig:fg_n5_cys}
    \end{subfigure}
    \hfill
    \begin{subfigure}[b]{0.44\textwidth}
        \centering
        \includegraphics[width=\textwidth]{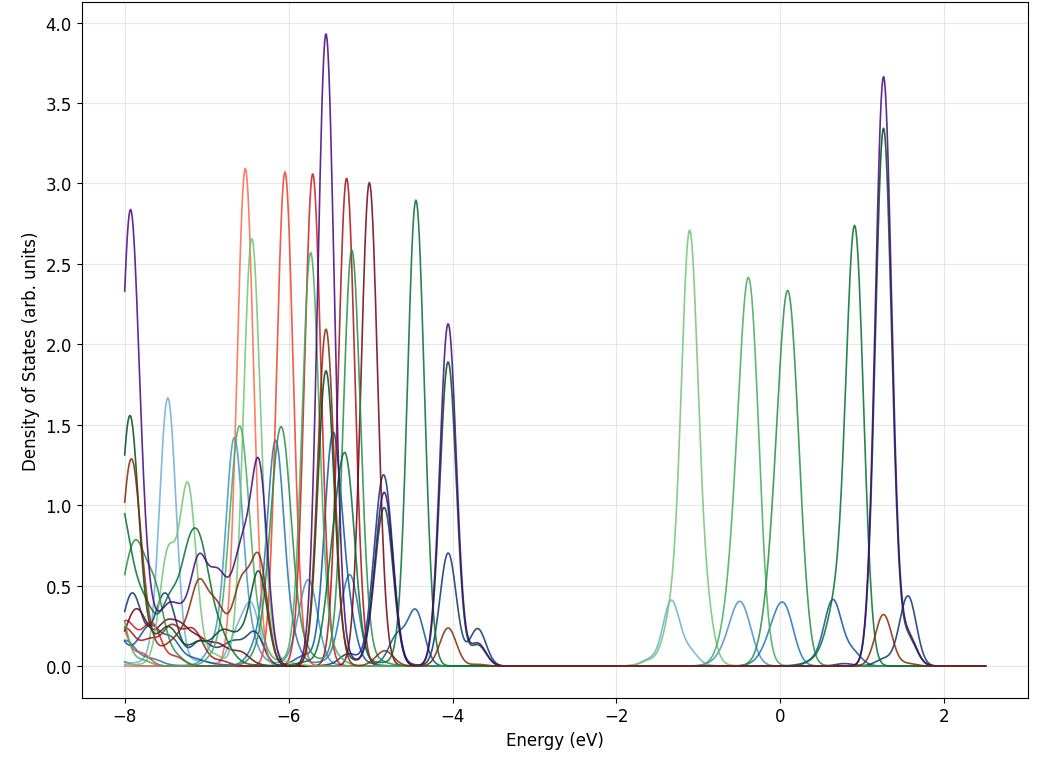}
        \caption{Lysine functional groups}
        \label{fig:fg_n5_lys}
    \end{subfigure}
    \caption{
        Functional group-projected PDOS for C[AAAAK]$_5$. \textbf{(a)} cysteine and first alanine (Ala1) and \textbf{(b)} all the lysine residues. }
    \label{fig:si_fg_n5}
\end{figure}

Having established motif robustness, we now identify the specific chemical moieties governing frontier orbital states. Figures~\ref{fig:si_fg_n1} and \ref{fig:si_fg_n5} decompose the PDOS of C[AAAAK] and C[AAAAK]$_5$ by functional groups within cysteine (Cys), lysine (Lys) residues. Critically, carbonyl (C=O) groups—primarily dominate the frontier orbitals in both structures.
In Fig. ~\ref{fig:si_fg_n5}, we have shown the decomposition into functional groups for all the lysines, confirming that carbonyl groups are the major contributors in all the constituent amino acids.

\FloatBarrier
\section{Geometrical response of C[AAAAK]$_3$ to applied electric field}
\label{sec:geom_response}

To characterize the structural perturbation induced by an external electric field, we measured C=O bond lengths and their projections onto the helix axis with point the charges of the magnitude 0.3 q/e and with different signs (positve at C terminus and negative at the N terminus) at 20~\AA\ from the helix center (field application point). Table~\ref{tab:field_geometry} quantifies these changes. One can see the small differences toward the C-terminus.

\begin{table}[htbp]
\centering
\small
\caption{Effect of an applied electric field on the C=O
bond lengths and their projections along the helix axis
in C[AAAAK]$_3$. The differences are calculated as
$\Delta = \mathrm{After} - \mathrm{Before}$. All values are given in \AA.}
\label{tab:field_geometry}

\setlength{\tabcolsep}{4pt}
\renewcommand{\arraystretch}{1.1}

\begin{tabular}{lrrrrrr}
\toprule
& \multicolumn{3}{c}{C=O bond length (\AA)}
& \multicolumn{3}{c}{Axial projection (\AA)} \\
\cmidrule(lr){2-4}
\cmidrule(lr){5-7}
\textbf{Residue}
& \textbf{Before}
& \textbf{After}
& $\boldsymbol{\Delta}$
& \textbf{Before}
& \textbf{After}
& $\boldsymbol{\Delta}$ \\
\midrule
Cysteine   & 1.240 & 1.242 &  0.001 & 1.172 & 1.180 & +0.009 \\
Alanine 1  & 1.234 & 1.234 &  0.000 & 1.184 & 1.182 & -0.002 \\
Alanine 2  & 1.235 & 1.236 &  0.001 & 1.154 & 1.158 & +0.004 \\
Alanine 3  & 1.237 & 1.238 &  0.000 & 1.186 & 1.190 & +0.004 \\
Alanine 4  & 1.237 & 1.237 &  0.000 & 1.172 & 1.175 & +0.003 \\
Lysine 1   & 1.239 & 1.240 &  0.001 & 1.188 & 1.189 & +0.001 \\
Alanine 5  & 1.239 & 1.240 &  0.001 & 1.189 & 1.191 & +0.002 \\
Alanine 6  & 1.239 & 1.240 &  0.001 & 1.184 & 1.186 & +0.002 \\
Alanine 7  & 1.239 & 1.241 &  0.001 & 1.174 & 1.179 & +0.005 \\
Alanine 8  & 1.239 & 1.239 &  0.000 & 1.173 & 1.177 & +0.004 \\
Lysine 2   & 1.238 & 1.239 &  0.000 & 1.186 & 1.187 & +0.001 \\
Alanine 9  & 1.237 & 1.238 & 0.002 & 1.187 & 1.189 & +0.002 \\
Alanine 10 & 1.241 & 1.242 & 0.001 & 1.186 & 1.190 & +0.004 \\
Alanine 11 & 1.230 & 1.231 & 0.001 & 1.131 & 1.137 & +0.007 \\
Alanine 12 & 1.230 & 1.231 & 0.001 & 1.071 & 1.099 & +0.028 \\
Lysine 3   & 1.216 & 1.218 & 0.002 & 1.123 & 1.143 & +0.019 \\
\bottomrule
\end{tabular}
\end{table}

\end{document}